\documentclass{pramana}

\usepackage[T1]{fontenc}
\usepackage{colortbl, xcolor 
}
\usepackage{amsmath}
\usepackage{hyphenat}
\usepackage{graphicx,color}
\usepackage{epstopdf}
\usepackage{tikz}
\usetikzlibrary{positioning}
\usepackage{graphicx,amsmath,bm}

\usepackage{hyperref}
\hypersetup{
    colorlinks=true,
    linkcolor=blue,
    filecolor=blue,      
    urlcolor=blue,
    citecolor=blue
}
\usepackage{caption}
\usepackage{subcaption}
\usepackage[noadjust]{cite}
\usepackage{titlesec}
\definecolor{Gray}{gray}{0.93}
\definecolor{Gray1}{gray}{0.95}
\definecolor{LightCyan}{rgb}{0.88,1,1}
\definecolor{mycolor}{rgb}{1,0.88,1}
\begin{document}

\title{Phenomenological study of $Z'$ in the minimal B$-$L model at LHC}

\author{Balasubramaniam. K. M.\textsuperscript{1, 2, *}}
\affilOne{\textsuperscript{1} Department of Physics $\&$ Astronomy, University of Sussex, UK\\}
\affilTwo{\textsuperscript{2} Theory Division, Physical Research Laboratory (PRL), India}


\twocolumn[{

\maketitle
\corres{kmbala86@gmail.com}


\msinfo{submitted on 6th Oct '16, for Pheno01@IISERM's Proceedings in Pramana - Journal of Physics.} 


\begin{abstract}
The phenomenological study of neutral heavy gauge boson ($Z^{\prime}_{B-L}$) of the minimal B-L extension was done on the dimuon production channel of the LHC.  The study begins with the LEP-II constraints on $Z'$ searches, and the dimuon events are simulated at the parton level at the CM energies of 7 TeV and 8 TeV and studied with an integrated luminosity of 1.21 $fb^{-1}$ and 20.5 $fb^{-1}$ respectively.  Later, the ATLAS detector-specific cuts unique to the Muon Pairs are imposed followed by the signal-selection-cuts on the Invariant Mass of the dimuon which restrict the events that are to be passed for Signal-Background Analysis, that are finally compared with the ATLAS data, and accounted for no experimental detection of $Z^{\prime}_{B-L}$ boson.  It has been simulated further at the CM energy of 14 TeV with an integrated luminosity of 300 $fb^{-1}$ to predict a possible discovery of this B-L neutral-heavy gauge boson with a mass corresponding to 1.5 TeV and a $Z'$ coupling strength of 0.2 based on the signal-background analysis.
\end{abstract}


}]



 
\pgrange{1-9}
\setcounter{page}{1}
\lp{9}

\section{Introduction}
The search on $Z'_{B-L}$ is done on the dimuon production channel of the p-p LHC collisions,  at the partonic level.  The Drell Yan process, $p \ p \ \rightarrow \ {\mu}^+ \ {\mu}^-$ (upto tree level) are simulated in the B-L model, (intermediated by $\gamma$, Z, $Z'_{B-L}$,  $h_1$ \footnote{\label{notec} SM Higgs} $\&$ $h_2$ \footnote{\label{noted} $B-L$ Model Higgs}) form the Signal-plus-Background, whereas in the SM process (intermediated by $\gamma$, Z $\&$ h \textsuperscript{\ref{notec}}) forms the Background-alone. We have done the signal-background analysis in studying the potential for $Z'_{B-L}$ discovery in the Large Hadron Collider (LHC). 

\subsection{Outline of the study}
 A brief introduction to the minimal B-L model is given in the following section (\ref{theory}) covering the details on the $B-L$ Lagrangian (\ref{lag}) and the spontaneous breaking of the B-L symmetry with the gauge boson spectrum (\ref{sym}).  In the third section (\ref{analysis}), the mechanics of this study are explained: starting from the Phenomenological tools (\ref{phenotools}) to the study of $Z'_{B-L}$ confidence level (CL) of a possible case (\ref{possiblecase}) and ruling out the sisters' categories of the reviewed case by comparing with the ATLAS experimental bounds (\ref{impcase}).  Lastly, we conclude by making remarks on the studied cases (\ref{conc}) and catch a glimpse of other experimentally ruled out cases (Figs.~\ref{fig:B} -~\ref{fig:F}) that are involved in this research.

\section{Theoretical Framework}\label{theory}
\subsection{The $B-L$ Model}
The B-L model is a triply-minimal extension of the Standard Model (SM) in the gauge, scalar and fermion sectors.  As the gauge sector of the SM is extended by a single U(1) factor related to Baryon minus Lepton (B-L) number, the $B-L$ gauge sector becomes minimal.  Similarly, the requirement of a complex scalar singlet for Spontaneous Breaking of $B-L$ Symmetry makes the scalar sector as minimal.  Thirdly, the introduction of an SM-singlet Right-Handed (RH) fermion per generation to eliminate the triangular $B-L$ gauge anomalies makes the fermion sector minimally extended.   The B-L charge is chosen to cure the new gauge and mixed U(1) gravitational anomalies.

\subsection{The Lagrangian of the  minimal $B-L$ model}\label{lag}
The Lagrangian of the minimal B-L Model obeying the $SU(3)_C \otimes SU(2)_L \otimes U(1)_Y \otimes U(1)_{B-L}$ gauge symmetry can be decomposed as:
\begin{eqnarray}
\centering
\mathcal{L} &=& \mathcal{L}_S + \mathcal{L}_{YM} + \mathcal{L}_f + \mathcal{L}_Y 
\end{eqnarray}
where the terms on the RHS are the scalar, Yang-Mills (YM)/gauge, fermion, and Yukawa parts respectively.

\subsubsection{Scalar sector}\label{scalar}\textcolor{white}{xyz}\\ \\
For the spontaneous breaking of B-L symmetry of the extra U(1) gauge group, a complex scalar singlet ($\chi$) is introduced along with the SM scalar doublet ($\Phi$).  Thus the scalar Lagrangian becomes,
 \begin{eqnarray*}
\centering
\mathcal{L}_S &=& (D^{\mu} {\Phi})^{\dagger}(D_{\mu}{\Phi}) + (D^{\mu}{\chi})^{\dagger}(D_{\mu}{\chi}) - V({\Phi}, \chi) \hskip 1.0cm
\end{eqnarray*}
with the scalar potential given by 
\begin{eqnarray*}
V({\Phi}, \chi) &=& m^2 {\Phi}^{\dagger}{\Phi} + {\mu}^2 |{\chi}|^2 +\\  \textcolor{white}{V({\Phi}, \chi)} && \left( \begin{array}{c} {\Phi}^{\dagger}{\Phi} \ \ |{\chi}|^2   \end{array} \right) \left( \begin{array}{cc} {\lambda}_1 & \frac{{\lambda}_3}{2} \\ \frac{{\lambda}_3}{2} & {\lambda}_2 \end{array} \right) \left( \begin{array}{c} {\Phi}^{\dagger}{\Phi} \\ |{\chi}|^2  \end{array} \right) \\ 
\textcolor{white}{V({\Phi}, \chi)} &=& m^2 {\Phi}^{\dagger}{\Phi} + {\mu}^2 |{\chi}|^2 + {\lambda}_1 ({\Phi}^{\dagger}{\Phi})^2 + \\ 
\textcolor{white}{V({\Phi}, \chi)} && {\lambda}_2 |{\chi}|^4 +  {\lambda}_3 {\Phi}^{\dagger}{\Phi} |{\chi}|^2
\end{eqnarray*}
where ${\Phi}$ and ${\chi}$ are the complex scalar Higgs doublet and singlet fields.  For ${\Phi}$ and ${\chi}$ fields, the $B-L$ charges are taken as 0 and +2 respectively. The charge of the ${\chi}$ field has been chosen to ensure the gauge invariance of the fermions sector of the minimal $B-L$ model. 

\subsubsection{Yang-Mills / Gauge sector}\textcolor{white}{xyz}\\ \\
The non-Abelian field strengths of this model are the same as in the SM whereas the Abelian ones can be written as follows:
\begin{eqnarray*}
\centering
\mathcal{L}_{YM}^{Abel} &=& - \frac{1}{4} F^{{\mu}{\nu}}F_{{\mu}{\nu}} - \frac{1}{4} F^{'{\mu}{\nu}}F^{'}_{{\mu}{\nu}},
\end{eqnarray*}
\begin{eqnarray*}
where, \  
F_{{\mu}{\nu}} &=& {\partial}_{\mu}B_{\nu} - {\partial}_{\nu}B_{\mu}, \\
F^{'}_{{\mu}{\nu}} &=& {\partial}_{\mu}B'_{\nu} - {\partial}_{\nu}B'_{\mu}
\end{eqnarray*}
The fields $ B_{\mu}$ and $B'_{\nu} $ are the $U(1)_Y$ and $U(1)_{B-L}$ gauge fields respectively.  In this field basis, the covariant derivative is 
\begin{eqnarray*}
\centering
D_{\mu}   &\equiv&  {\partial}_{\mu} + ig_S T^{\alpha}G^{{}{\alpha}}_{\mu} + igT^{\alpha}W_{\mu}^{\alpha}\\ 
\textcolor{white}{D_{\mu}} && +\ ig_1 Y B{\mu} + i( \tilde{g}Y + g'_1 Y_{B-L} )B'_{\mu}
\end{eqnarray*}
The gauge couplings $\tilde{g}$ and $ g'_1$ are free parameters.   The pure $B-L$ model is defined by the condition $\tilde{g} = 0$ (i.e., the free parameter $\tilde{g}$ is nullified at the EW scale).  This implies no mixing at the tree level between the $ Z'$ bosons of $B-L$ Model and the Z bosons of the SM.

\subsubsection{Fermion sector}\textcolor{white}{xyz}\\ \\
The fermion Lagrangian density (with k being the generation index) is given by 
\begin{eqnarray*}
\centering
\mathcal{L}_f &=&  \sum_{k=1}^3   \Big(\ i{\overline{q}}_{kL}{\gamma}_{\mu}D^{\mu}q_{kL}\  +\ i{\overline{q}}_{kR}{\gamma}_{\mu}D^{\mu}q_{kR}\ \ +\\ 
 && \textcolor{white}{abcde}i{\overline{d}}_{kR}{\gamma}_{\mu}D^{\mu}d_{kR}\  \ + \ \ i{\overline{l}}_{kL}{\gamma}_{\mu}D^{\mu}l_{kL}\ \ + \\
  && \textcolor{white}{abcde}i{\overline{e}}_{kR}{\gamma}_{\mu}D^{\mu}e_{kR}\ \ + \ \ i{\overline{\nu}}_{kR}{\gamma}_{\mu}D^{\mu}{\nu}_{kR} \
\Big ) 
\end{eqnarray*}
 
where the fields' charges are the usual SM and $B-L$ ones (in particular, $B-L  = \frac{1}{3}$ for quarks and -1 for leptons with no distinction between generations, hence ensuring universality).  The $B-L$ charge assignments of the fields as well as the introduction of new fermion RH heavy neutrinos (${\nu}_{R}$'s, charged -1 under $B-L$) are designed to eliminate the triangular $B-L$ gauge anomalies of the theory. \\
\indent Therefore, the $B-L$ gauge extension of the SM gauge group broken at the TeV scale necessarily requires at least one new scalar field and three new fermion fields which are charged with respect to the $B-L$ group.

\subsubsection{Yukawa sector}\label{yuk}\textcolor{white}{xyz}\\ \\
Finally, the Yukawa interactions are 
\begin{eqnarray*}
\centering
\mathcal{L}_Y &=& \sum_{i,j,k=1}^3  
-\ y^d_{jk} {\overline{q}}_{jL}d_{kR}\Phi\ -\ 
y^u_{jk} {\overline{q}}_{jL}u_{kR}\tilde{\Phi} \\
 &&  \textcolor{white}{abcd}\ -\ y^e_{jk} {\overline{l}}_{jL}e_{kR}\Phi \ - \  y^{\nu}_{jk} {\overline{l}}_{jL} {\nu}_{kR}\tilde{\Phi} \\
  &&  \textcolor{white}{abcd}\ -\ y^M_{jk} ({\overline{\nu}_{R}})^c_j{\nu}_{kR} {\chi}\ +\ h.c. 
\end{eqnarray*}
where  $ \tilde{\Phi} = i{\sigma}^2 {\Phi}^*  $ and the last term is the Majorana contribution, and the others are the usual Dirac ones.  While working on the basis in which the RH neutrino Yukawa coupling matrices, $y^M$ are diagonal, real, and positive, these are the only allowed gauge invariant terms.  The last term in the above equation combines the neutrinos to the new scalar singlet field, $\chi$, which allows the dynamical generation of neutrino masses, and acquires a VEV through the Higgs mechanism.
 
\subsection{Spontaneous breaking of $B-L$ Symmetry $\&$ gauge boson spectrum}\label{sym}
In the Feynman gauge, the scalar fields ( $ \Phi \ \& \ \chi $ ) can be parametrized  \cite{Basso:2011hn} as  
\begin{eqnarray*}
\centering
\Phi &=& \frac{1}{2} \left( \begin{array}{c} -i({\omega}^1 - i{\omega}^2) \\ {\nu} + (h+i{z})   \end{array} \right) \\ \\
{\chi} &=& \frac{1}{\sqrt{2}} (x + (h' + iz'))
\end{eqnarray*}
where ${\omega}^{\pm} = {\omega}^1 {\mp} i{\omega}^2,\ z\ \&\ z' $ are the would-be Goldstone bosons of $W^{\pm}$, Z and Z' respectively.   
\begin{eqnarray*}
\centering
\textcolor{black}{D^{\mu} {\Phi} ( D_{\mu} {\Phi} )^{\dag}} &=& \frac{1}{2} ({\partial}^{\mu}h) ({\partial}_{\mu}h) + \frac{1}{8} (h + {\nu})^2\   \Big [\ g^2\ |\ W_1^{{}{\mu}}\\ \\
\textcolor{white}{D^{\mu} {\Phi} ( D_{\mu} {\Phi} )^{\dag}} && - iW_2^{{}{\mu}}\ |^2 + (\ gW_3^{{}{\mu}} - g_1B^{\mu} - \tilde{g}B'^{\mu}\ )^2\ \Big] 
\end{eqnarray*}
and
\begin{eqnarray*}
D^{\mu}{\chi}(D_{\mu}{\chi})^{\dagger} &=& \frac{1}{2}  ({\partial}^{\mu}h') ({\partial}_{\mu}h') + \frac{1}{2}(h' + x)^2\ (\ g'_1 2 B'^{\mu} \ )^2
\end{eqnarray*}
where we have taken $ Y_{\chi}^{B-L} = 2 $ to guarantee the gauge invariance of the Yukawa terms (\ref{yuk}).  In the first of the above two equations, we can recognise immediately the SM charged gauge bosons $W^{\pm}$, with $ M_W = \frac{g {\nu}}{2} $ as in the SM.  The other gauge boson masses are not so simple to identify, because of mixing.  In analogy with the SM, the fields of definite mass are linear combinations of $ B^{\mu},\ W_3^{\mu}  $ and $ B'^{\mu} $.  The explicit expressions are: 
\begin{eqnarray*}
\left( \begin{array}{ccc}  B^{\mu}  \\W_3^{\mu}  \\  B'^{\mu}  \end{array} \right) &=& X \hskip 0.5cm  \left( \begin{array}{ccc}  A^{\mu}  \\Z^{\mu}  \\  Z'^{\mu}  \end{array} \right)    
\end{eqnarray*}
where,
\begin{eqnarray*}
X  &=&  
\left( \begin{array}{ccc}cos{\vartheta}_{\omega}  \hskip0.3cm  -sin {\vartheta}_{\omega}cos {\vartheta}'   \hskip1.0cm  sin {\vartheta}_{\omega} sin{\vartheta}'  \\
sin {\vartheta}_{\omega} \hskip0.9cm cos {\vartheta}_{\omega} cos{\vartheta}' \hskip0.4cm  -cos {\vartheta}_{\omega}sin {\vartheta}' \\
0 \hskip2.4cm sin {\vartheta}' \hskip1.8cm  cos {\vartheta}'
\end{array} \right)
\end{eqnarray*}
with $ - \frac{\pi}{4} \leq {\vartheta}' \leq \frac{\pi}{4} $, such that:
\begin{eqnarray*}
\centering
tan 2 {\vartheta}' &=& \frac{2 \tilde{g} \sqrt{g^2 + g_1^2}}{\tilde{g}^2 + 16 (\frac{x}{\nu})^2 g_1^{'2} - g^2 - g_1^2} 
\end{eqnarray*}
The gauge boson masses are:
\begin{eqnarray}
\centering
M_A &=& 0
\end{eqnarray}
 
Now, setting $\tilde{g}$ to 0 for the pure $B-L$ model, the mixing angle ${\vartheta}'$ vanishes, implying no mixing, at the tree level, between the $Z_{SM}$ and $Z'_{B-L}$ bosons.  The $Z$ and $Z'_{B-L}$ masses are:
\begin{eqnarray}
\centering
M_Z &=& \sqrt{g^2 + g_1^2}.\frac{\nu}{2},\\
M_{Z'_{B-L}} &=& 2g'_1 x
\end{eqnarray}
 
\indent To complement the section on the $B-L$  model, we summarise the mass eigenstates and the assignation of Hypercharge (Y) and B-L quantum number ($B-L$) to the chiral fermionic and scalar fields in Tables ~\ref{tab:mass eigenstates} $\&$ ~\ref{tab:B-L-quantum numbers} respectively.

\begin{table}[]
\centering
\renewcommand{\arraystretch}{2.0}
\begin{tabular}{|p{1.0cm} || p{1.0cm} | p{1.2cm} || p{1.4cm} || p{1.4cm} |} \hline 
Name & Quarks & Leptons & Neutrinos & Higgses \\ \hline
 \textbf{${\psi}$} & q & l & ${\nu}_l \  \& \  {\nu}_h$ & $h_1 \ \& \ h_2$ \\  
\textbf{${Mass}$} & $m_q$ & $m_l$ & $m_{{\nu}_l} \ \& \ m_{{\nu}_h}$ & $m_{h_1}\  \&  \ m_{h_2}$ \\ \hline
\end{tabular}
\caption{Mass eigenstates \cite{Basso:2011hn}}
\label{tab:mass eigenstates}
\end{table}
 
\begin{table}[h]
\centering
\renewcommand{\arraystretch}{ 2.5}
 \begin{tabular}{| p{1.2cm} || c | c | c || c | c | c || c | c |} \hline
 \textbf{ ${\psi}$}    &   $q_L$   &   $u_R$    &  $d_R$    &   $l_L$   &   $e_R$   &   ${\nu}_R$   &   $\Phi$    &    ${\chi}$ \\  \hline \hline 
 \textbf{${SU(3)_C}$} & 3 & 3 & 3 & 1 & 1 & 1 & 1 & 1 \\     
 \textbf{${SU(2)_L}$} & 2 & 1 & 1 & 2 & 1 & 1 & 2 & 1 \\ 
 \textbf{${Y}$}  & $\frac{1}{6}$ & $\frac{2}{3}$ & $ - \frac{1}{3}$ & $- \frac{1}{2}$ & $-1$ & $0$ & $\frac{1}{2}$ & $0$ \\ 
 \textbf{${B-L}$} & $\frac{1}{3}$ & $\frac{1}{3}$ & $\frac{1}{3}$ & -1 & -1 & -1 & 0 & 2 \\ \hline
 \end{tabular}
\caption{Hypercharge ($Y$) and $B-L$ quantum number assignation to chiral fermion and scalar fields \cite{Basso:2011hn}.}
\label{tab:B-L-quantum numbers}
\end{table}
 
\section{$Z'_{B-L}$ - Analysis $\&$ Results}\label{analysis}

\subsection{Phenomenological Tools}\label{phenotools}
\begin{figure}
\begin{subfigure}{0.5\textwidth}
\begin{tikzpicture}
\hskip1.0cm \node (img1)  {\includegraphics[scale=0.5, trim={6.5cm 14.5cm 0 4cm}, clip]{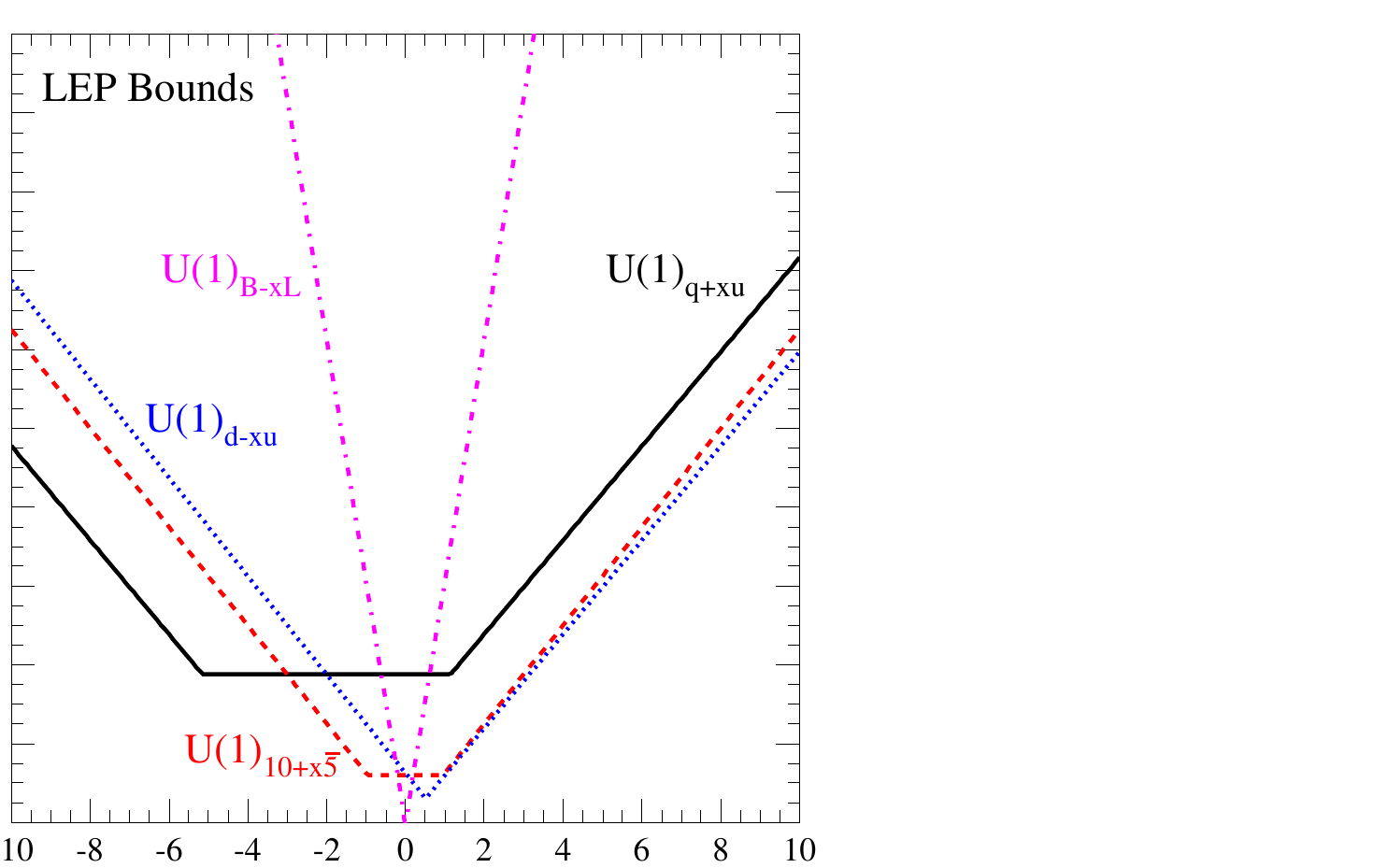}};
  \node[below=of img1, node distance=0cm, yshift=1cm,font=\color{black}] {x};
  \node[right=of img1, node distance=0cm, rotate=90, anchor=center,yshift=9.1cm,font=\color{black}] {$M_{Z'}/g'_1$ (in TeV)};
\end{tikzpicture}
\end{subfigure}
\caption {LEP-II constraints \cite{Carena:2004xs} on $\frac{M_{Z'}}{g'_1}$ of various Z' models.  For the value of x=1 (i.e., for B-L Model), only $\frac{M_{Z'}}{g'_1}\ \geq\ $ 6 TeV are allowed.}
\label{fig:lep}
\end{figure}

The search for $Z'_{B-L}$ boson is done on the dimuon production channel of the LHC process, $p \ p\ \rightarrow {\mu}^+\ {\mu}^- $.  We have taken the FeynRules model file for the Minimal $B-L$ model, implemented by L. Basso and G. M. Pruna \cite{pureB-L}.  The motivation was to get myself introduced to the techniques of FeynRules \cite{Christensen:2008py}.  We used FeynRules2.0 \cite{Alloul:2013bka} and Mathematica (version - 9) \cite{wolf} to generate the Universal FeynRules Output (UFO) \cite{Degrande:2011ua} files which were fed into the event generator, MadGraph5.0 \cite{Alwall:2014hca}.  We have used the inbuilt Parton Density Function (PDF) set, CTEQ6L1 in MadGraph5.0 \cite{Alwall:2014hca} for our study.  We have resorted to the standard Large Electron Positron (LEP)-II bounds \cite{Carena:2004xs} on $Z'$ searches as shown in Fig.~\ref{fig:lep} to standardise the values of the parameters for our study.  

\subsection{Initialisation of $Z'_{B-L}$ Parameters}\label{parametersInit}
We have chosen the values of $Z'$ coupling constant ($g'_1$)  \footnote{\label{notea}Dataset-specific-parameters: Z-Prime's coupling constant ($g'_1$), mass ($M_{Z'}$), and total decay width (${\Gamma}_{Z'}$).} as 0.2 and 0.5 with four different mass ($M_{Z'}$  in TeV)\textsuperscript{\ref{notea}} parameters of values: 1.5, 2.0, 3.0, and 5.0 respectively in accordance with the bounds from the LEP-II experiment.  The expression for the partial decay widths ( ${\Gamma} _{\ Z' \rightarrow f \bar{f} }$ ) of $Z'_{B-L}$ into SM fermions has been derived from the coupling vertex of $Z'_{B-L}$ with the SM fermions is:
\begin{eqnarray*}
\centering
    {\Gamma}_{\ Z' \rightarrow f \bar{f} \ } &=& \frac{M_{Z'}}{12 \pi} \ C_{f}\ (v^f)^2\  \Big[\ {1+2\frac{m_{f}^2}{M_{Z'}^2}}\ \Big] \
   \sqrt{1-\frac{4m_{f}^2}{M_{Z'}^2}}
\end{eqnarray*}
where, $C_f$ is the color-factor of the fermion ($f$), ${\nu}^f$ ( = $(B-L)\ g'_1$), is the coupling between the $B-L$ charge of fermion ($f$) and $Z'$ coupling b/w SM fermions ($g'_1$) and $m_f$ is the mass of fermion ($f$).  The total decay width (${\Gamma}_{Z'}$)\textsuperscript{\ref{notea}} of $Z'_{B-L}$ boson is the sum of all the partial decay widths (${\Gamma}_{\ Z' \rightarrow f \bar{f} \ }$) of $Z'_{B-L}$ into SM-fermion pairs except the neutrino-pairs (taking the SM-neutrinos as massless). These set of parameters are grouped into six datasets as shown in Table ~\ref{tab:parameters}.

\begin{table}[h]
\begin{minipage}{0.5\textwidth} 
\centering
\begin{tabular}{| p{1.0cm} || p{1.0cm}| p{1.0cm} || p{1.0cm} |}
\hline
\rowcolor{Gray}
 	& 	             \textcolor{blue}   {0.2}		   &               \textcolor{blue}  {0.5}                          &                     \textcolor{blue}   {\textbf{ $g'_1$}} \\ \hline \hline
\rowcolor{Gray1}
\vskip0.2cm  \textcolor{violet}{1500} & \vskip 0.2cm  \textcolor{red}{38.20} \vskip 0.2cm \textcolor{darkgray}{\fbox{A}}   & \vskip 0.2cm  \textcolor{red}{nil \footnote{\label{noteb} The datasets corresponding to the parameters $g_1'$ = 0.5 with $M_{Z'} = 1.5 \ \& \ 2.0$ TeV are discarded as the LHE files corresponding to them got corrupted. } } \vskip 0.2cm    & \vskip 0.2cm    \textcolor{red}{\textbf{{${\Gamma}_{Z'}$  (in GeV)}}}\\ \hline
\rowcolor{Gray}
\vskip0.2cm \textcolor{violet}{2000} & \vskip 0.2cm  \textcolor{red}{50.93} \vskip 0.2cm  \textcolor{darkgray}{\fbox{B}}  & \vskip 0.2cm  \textcolor{red}{nil \textsuperscript{\ref{noteb}}}  \vskip 0.2cm    & \vskip 0.2cm   \textcolor{red}{{\textbf{${\Gamma}_{Z'}$ (in GeV)}}}\\ \hline
\rowcolor{Gray1}
\vskip0.2cm \textcolor{violet}{3000} & \vskip 0.2cm  \textcolor{red}{76.39} \vskip 0.2cm  \textcolor{darkgray}{\fbox{C}}  & \vskip 0.2cm  \textcolor{red}{477.45} \vskip 0.2cm  \textcolor{darkgray}{\fbox{D}}  &   \vskip 0.2cm   \textcolor{red}{\textbf{{${\Gamma}_{Z'}$ (in GeV)}}}\\ \hline
\rowcolor{Gray}
\vskip0.2cm \textcolor{violet}{5000} & \vskip 0.2cm  \textcolor{red}{127.32} \vskip 0.2cm  \textcolor{darkgray}{\fbox{E}}  & \vskip 0.2cm  \textcolor{red}{795.75} \vskip 0.2cm  \textcolor{darkgray}{\fbox{F}}  &  \vskip 0.2cm   \textcolor{red}{\textbf{${\Gamma}_{Z'}$ (in GeV)}}\\ \hline \hline
\rowcolor{Gray1}
\textcolor{violet}{\textbf{{$ M_{Z'} $ (in GeV)}}} &            &             &              \\ 
\hline
\end{tabular}
\vskip 0.75cm
where, \\ 
 \textcolor{darkgray}{\fbox{C}} : the Z' signal with  \textcolor{violet}{$M_{Z'}$ = 3000 GeV},  \textcolor{blue}{$g'_1$ = 0.2} and  \textcolor{red}{${\Gamma}_{Z'}$ = 76.39 GeV} \\ 
\caption  {The datasets with different values of 3 parameters viz., mass ($M_{Z'}$), coupling of $Z'_{B-L}$ with SM Fermions ($g'_1$) and the total decay width of $Z'_{B-L}$ into SM Fermions (${\Gamma}_{Z'}$) upon which this study                is based.}
\label{tab:parameters}
\end{minipage}
\end{table}
 
\indent Initially, the $B-L$ model (in the form of UFO \cite{Degrande:2011ua} files) has been imported in the MadGraph5.0's \cite{Alwall:2014hca} environment and the  process $p\ p\ \rightarrow \ {\mu}^+ \ {\mu}^-$ is generated.  Then, the dataset-specific-parameters \textsuperscript{\ref{notea}} are set in the created process directory to produce 1 million partonic-level events at the Centre of Mass (CM) energy of 7 TeV, per each Invariant-Mass (of charged lepton pairs) window of varied sizes.  These windows are to ensure no loss of any significant events across the whole range of simulation which has spanned across the entire Invariant Mass range of 0 - 7000 GeV of the final state Muon pairs.  The simulated events' format in compliant with the Les Houches Event Accord (LHEA) \cite{Alwall:2006yp} was generated as the Les Houches Event (LHE) \cite{Alwall:2006yp} files.

These event files (LHE) corresponding to each Invariant Mass window were fed into MadAnalysis5 \cite{Conte:2012fm} where they got joined into a single large LHE \cite{Alwall:2006yp} file.  Corresponding to a particular dataset, which encompasses the entire simulation range of width approximately 6000 - 7000 GeV of Muon Pairs' Invariant Mass.

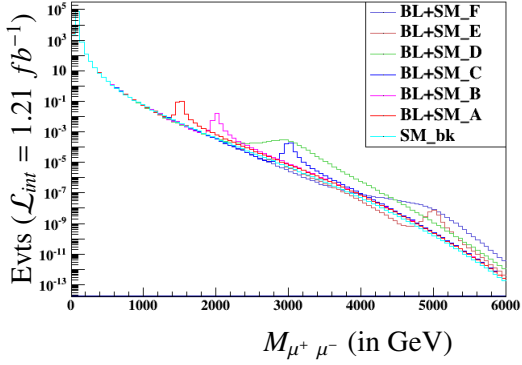
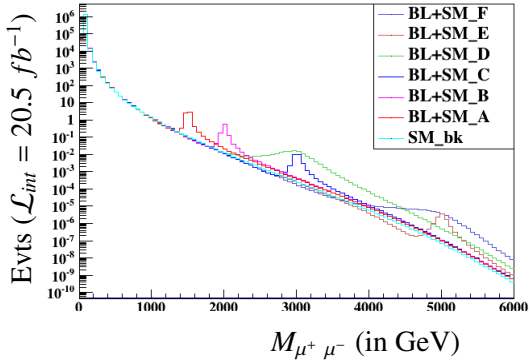
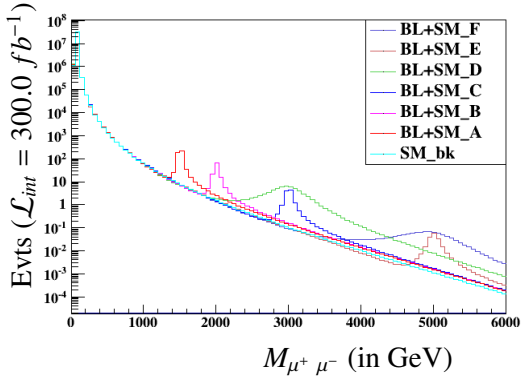
\begin{figure}[h!]
 
\begin{subfigure}{0.5\textwidth}
 
\begin{tikzpicture}
\hskip1.0cm \node (img1)  {\includegraphics[scale=0.5, trim={4.0cm 7.6cm 0 13.5cm}, clip]{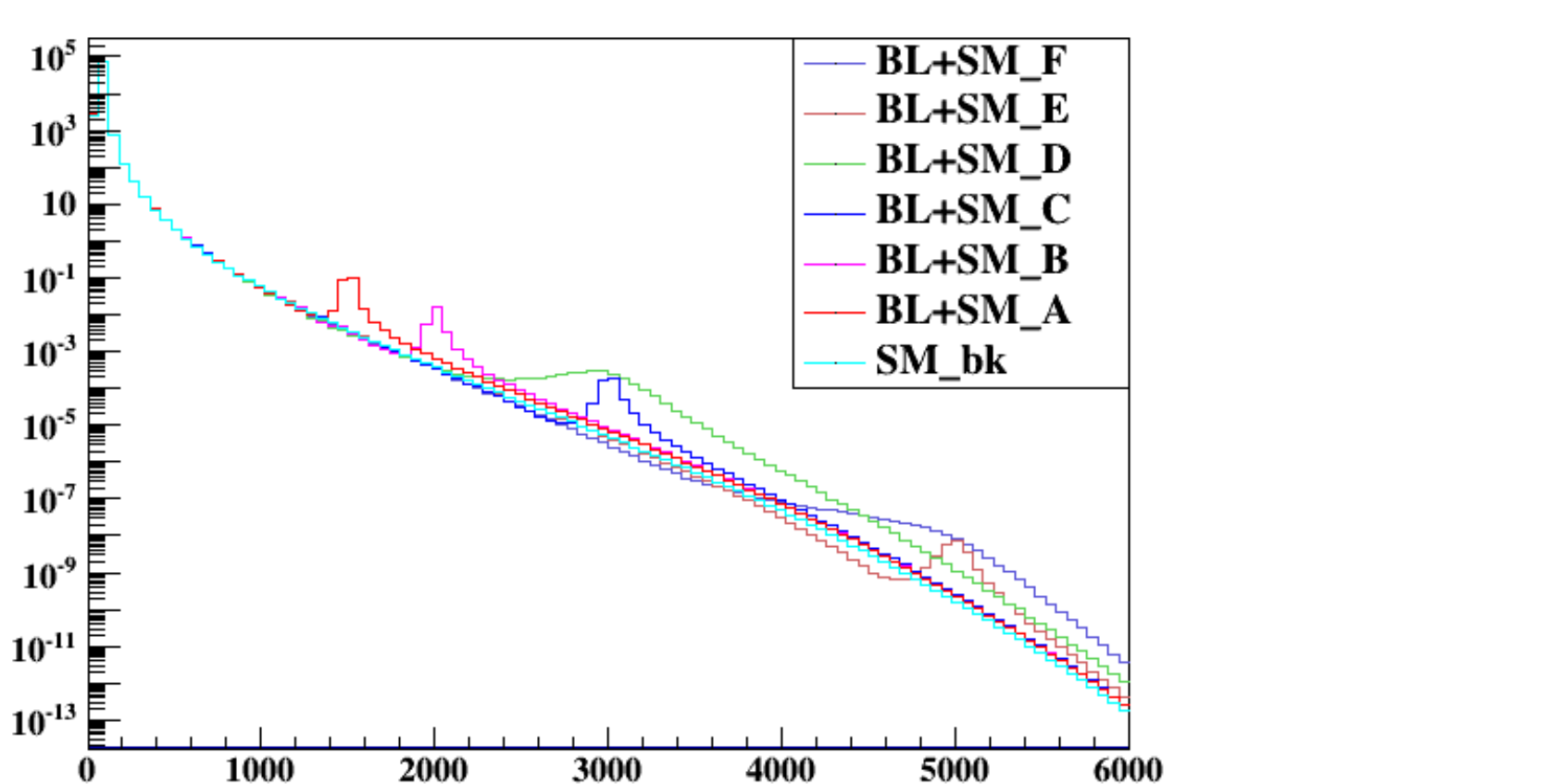}};
  \node[below=of img1, node distance=0cm, yshift=1cm,font=\color{black}] {$M_{{\mu}^+ \ {\mu}^-}$ (in GeV)};
  \node[right=of img1, node distance=0cm, rotate=90, anchor=center,yshift=9.8cm,font=\color{black}] {Evts ($\mathcal{L}_{int}$ = 1.21 $fb^{-1}$)};
\end{tikzpicture}
\caption{Simulated events at 7 TeV CM energies.}
\label{fig:7tev}
\end{subfigure}
\vskip0.4cm
\begin{subfigure}{0.5\textwidth}
\begin{tikzpicture}
\hskip1.0cm \node (img1){\includegraphics[scale=0.5, trim={4.0cm 17.6cm 0 3.5cm}, clip]{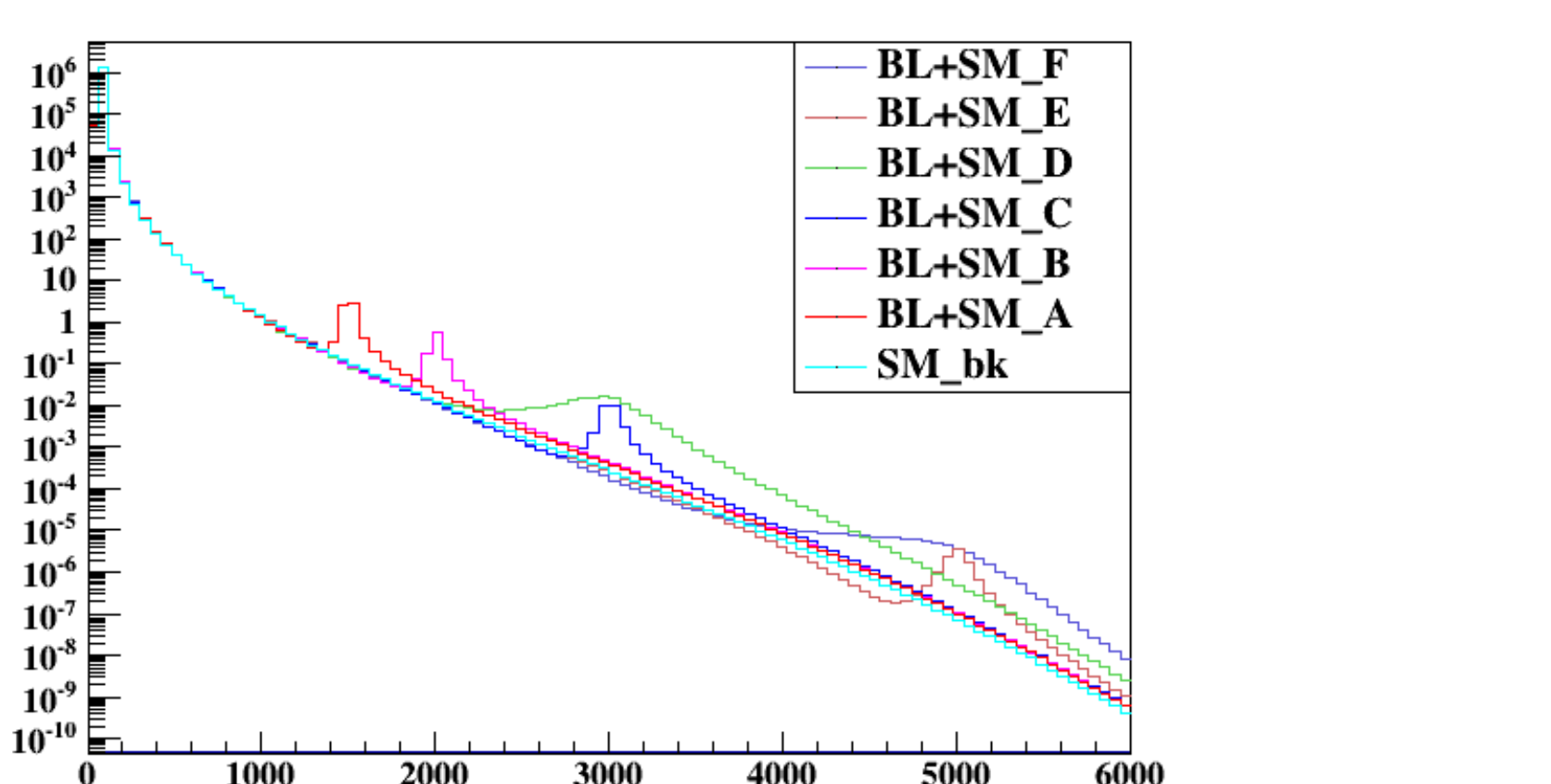}};
\node[below=of img1, node distance=0cm, yshift=1cm,font=\color{black}] {$M_{{\mu}^+ \ {\mu}^-}$ (in GeV)};
  \node[right=of img1, node distance=0cm, rotate=90, anchor=center,yshift=9.9cm,font=\color{black}] {Evts ($\mathcal{L}_{int}$ = 20.5 $fb^{-1}$)};
\end{tikzpicture}
\caption{Simulated events at 8 TeV CM energies.}
\label{fig:8tev}
\end{subfigure}
\begin{subfigure}{0.5\textwidth}
\begin{tikzpicture}
\hskip1.0cm \node (img1)  {\includegraphics[scale=0.5, trim={4.0cm 17.6cm 0 2.5cm}, clip]{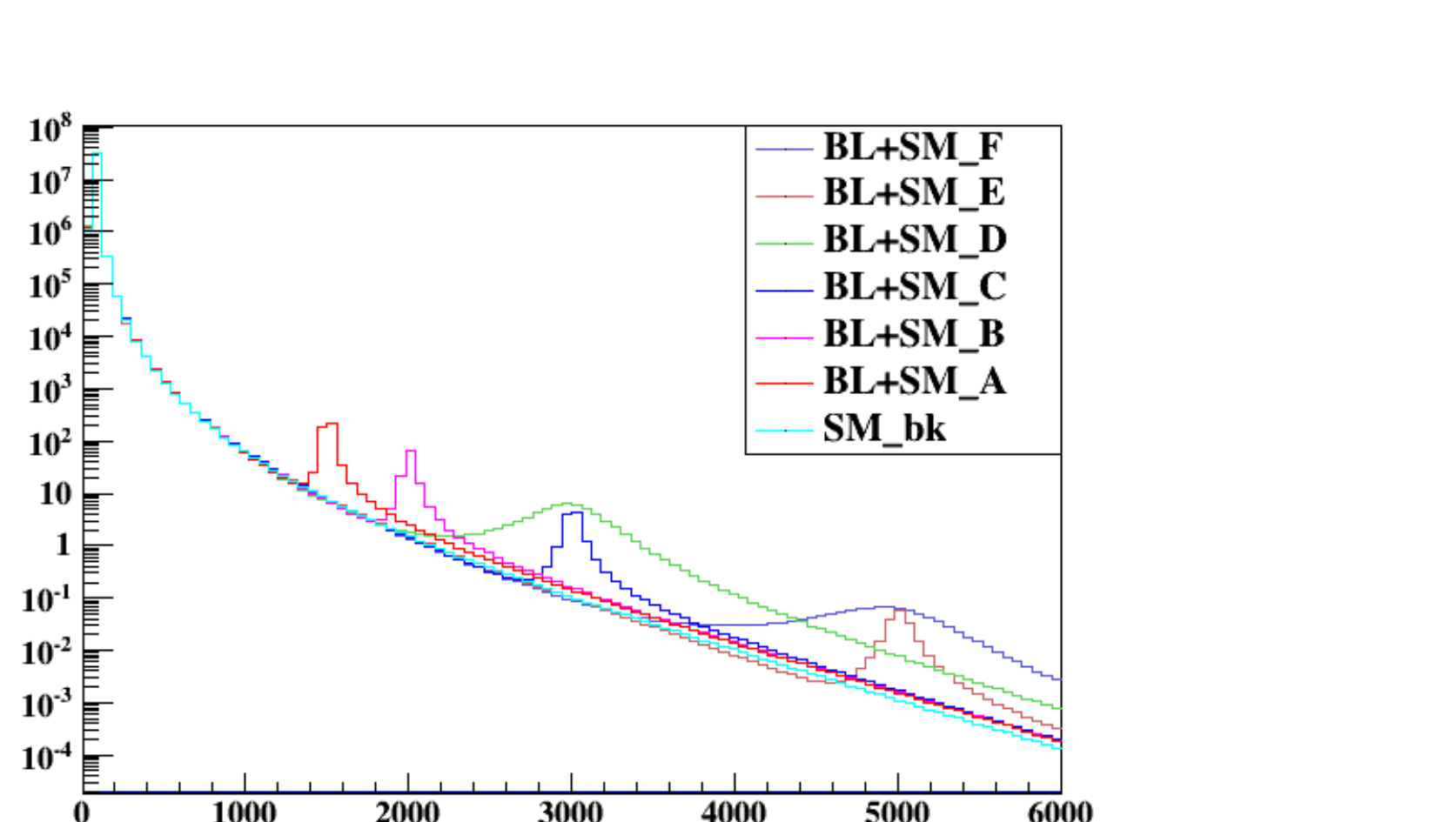}};
  \node[below=of img1, node distance=0cm, yshift=1cm,font=\color{black}] {$M_{{\mu}^+ \ {\mu}^-}$ (in GeV)};
  \node[right=of img1, node distance=0cm, rotate=90, anchor=center,yshift=9.8cm,font=\color{black}] {Evts ($\mathcal{L}_{int}$ = 300.0 $fb^{-1}$)};
\end{tikzpicture}

\caption{Simulated events at 14 TeV CM energies.}
\label{fig:14tev}
\end{subfigure}
\caption{$Z'_{B-L}$ Resonances in the process, $ p\ p\ \rightarrow \ {\mu}^+ \ {\mu}^-$ for considered datasets (signal + bkgrd) and dataset - SM (bkgrd alone) without any Kinematic Cuts.  The events-distributions of all the seven datasets are superimposed one upon other.  The resonance of the SM's Z boson at $M_{{\mu}^+ \ {\mu}^-}$ = 91.18 GeV can be seen very close to 0 GeV of $M_{{\mu}^+ \ {\mu}^-}$. }
\label{fig:all}
\end{figure}
 
\subsection{Comparison of 7, 8 $\&$ 14 TeV simulated events of all the datasets}\label{combdatasets}
At this juncture, the obtained single LHE file of massive size was normalised with an integrated luminosity of 1.21 $fb^{-1}$. The whole machinery of section (3.1) was repeated for the remaining datasets.  All of these datasets are an admixture of $Z'_{B-L}$ signal and the SM background.  For the analysis of $Z'_{B-L}$ signal from the background, another LHE file of large size has been generated as previously for the SM process.  This generation was done by importing the default model SM in the MG5 environment and simulated the events for the dimuon production from proton-proton collisions with the same PDF set, CTEQ6L1.  

The distribution of Events versus the Muon Pairs' Invariant mass is shown in Figure~\ref{fig:7tev} at the CM energy of 7 TeV with an integrated luminosity of ${\mathcal{L}}_{int}$ = 1.21 $fb^{-1}$.  We have repeated the similar study with the collision energy of 8 TeV and 14 TeV and normalised the generated partonic-level events with an integrated luminosity of  ${\mathcal{L}}_{int}$ = 20.5 $fb^{-1}$ and  ${\mathcal{L}}_{int}$ = 300.0 $fb^{-1}$ respectively.  The events distribution with the Muon Pairs' Invariant mass for the two cases just discussed is shown in Figures~\ref{fig:8tev} and~\ref{fig:14tev}.

We shall observe an increase in the total number of events as we move from the Events Distributions studied at the CM energies of 7 TeV to 8 TeV and then to 14 TeV respectively.  The events distributions corresponding to all the datasets are colour coded distinctly in all the detailed figures.  The signal peaks become prominent with increasing collision energy and integrated luminosity, which is very conspicuous in the Figures~\ref{fig:7tev},~\ref{fig:8tev} and~\ref{fig:14tev}. This variation in the signal's prominence shall be accounted for a potential discovery of $Z'_{B-L}$ in the upcoming sections.

The events distributions shown in Fig.~\ref{fig:all} have six $Z'_{B-L}$ signal peaks corresponding to the Signal-plus-Back ground-datasets, and one peak at $M_{{\mu}^+ \ {\mu}^-}$ = 91.18 GeV corresponding the SM Z-resonance where the six Signal-plus-Background datasets (viz., A, B, C, D, E $\&$ F) coincide with Background-Alone dataset (dataset - SM).

\subsection{Signal Vs Background Analysis for $Z'_{B-L}$ Resonance in Datasets - A $\&$ SM}
The LHE files of dataset - A and the dataset - SM generated at 7 TeV CM energy were fed into MadAnalysis5 \cite{Conte:2012fm} environment and normalised to an integrated luminosity of  ${\mathcal{L}}_{int}$ = 1.21 $fb^{-1}$.  

Then, two detector-specific-kinematic-cuts \footnote{\label{note2} (ATLAS) Detector-specific-kinematic-cuts on final state Muon Pairs: $P_T$ > 20.0 GeV and $\eta$ < 2.4 (common to all the datasets).} were applied on the transverse momenta ($P_T$ > 20.0 GeV) \textsuperscript{\ref{note2}} and Pseudorapidity  ($\eta$ < 2.4) \textsuperscript{\ref{note2}} of the final state Muon Pairs and the events satisfying these cuts were selected for further study.  We have calculated the cumulative efficiency (which is a ratio between the selected events and the sum of selected $\&$ rejected events) after the cuts and checked that it's never greater than 1, to ensure that we didn't lose any significant events with these cuts.  
 
\indent The signal-specific-kinematic-cuts \footnote{\label{note3} Signal-specific-kinematic-cuts on $M_{{\mu}^+ \ {\mu}^-}$ of Datasets (with Sgnl+Bkgrd) : \newline $\textcolor{white}{x}692.7\ GeV\ \leq \ M_{{\mu}^+ \ {\mu}^-} \ \leq \ 2307.3\ GeV \ ( \ for \ {\Gamma}_{SW}^{signl-A} \ )$ \newline $ \textcolor{white}{x}923.6\ GeV\ \leq \ M_{{\mu}^+ \ {\mu}^-} \ \leq \ 3076.4\ GeV \ ( \ for \ {\Gamma}_{SW}^{signl-B} \ ) $ \newline $1385.4\ GeV\ \leq \ M_{{\mu}^+ \ {\mu}^-} \ \leq \ 4614.6\ GeV \ ( \ for \ {\Gamma}_{SW}^{signl-C} \ ) $\newline$ \textcolor{white}{x}783.8\ GeV\ \leq \ M_{{\mu}^+ \ {\mu}^-} \ \leq \ 5216.2\ GeV \ ( \ for \ {\Gamma}_{SW}^{signl-D} \ )$ \newline $ 2309.0\ GeV\ \leq \ M_{{\mu}^+ \ {\mu}^-} \ \leq \ 7691.0\ GeV \ ( \ for \ {\Gamma}_{SW}^{signl-E} \ )$\newline$ 1306.4\ GeV\ \leq \ M_{{\mu}^+ \ {\mu}^-} \ \leq \ 8693.6\ GeV \ ( \ for \ {\Gamma}_{SW}^{signl-F} \ )$ \newline \textcolor{white}{abc} (same for datasets at 7, 8 $\&$ 14 TeV collision energies.)  } (common for studies done at 7, 8 $\&$ 14 TeV CM energies.) on the Invariant Mass ($M_{{\mu}^+ \ {\mu}^-}$) have been chosen as a Sampling Window (SW) with width (${\Gamma}_{SW}$) is:
\begin{eqnarray}
 {\Gamma}_{SW}^{signl-A} &=& M_{Z'}^{signl-A}\ \pm 3 \ {\Gamma}_{Z'}^{sgnl-A \textsuperscript{\ref{note3}}} 
\end{eqnarray}
\indent Thus, for dataset - A, with $M_{Z'}^{signl-A}$ = 1500 GeV $\&$ ${\Gamma}_{Z'}^{sgnl-A}$ = 38.20 GeV, the signal-specific-cuts can be implemented by selecting the events whose invariant mass falls within the range,
\begin{eqnarray*}
\centering
692.7\ GeV\ \leq \ M_{{\mu}^+ \ {\mu}^-} \ \leq \ 2307.3\ GeV \ ( \ for \ {\Gamma}_{SW}^{signl-A} \ ) \ \textsuperscript{\ref{note3}}
\end{eqnarray*}
\indent The event distribution with the Muon Pairs' invariant mass for dataset - A, at 7 TeV Collision energy is shown in Figure~\ref{fig:A7}.  By repeating the formerly mentioned procedures, the event distributions for dataset - A, at 8 TeV $\&$ 14 TeV were obtained which are shown in Fig.~\ref{fig:A8a} and Fig.~\ref{fig:A14} respectively.  \\
\indent The Confidence Level (CL) / statistical significance of the $Z'_{B-L}$ signal has been calculated using:
\begin{eqnarray*}
K_{Signal} &=& K_A - K_{SM} \\
Signal's \ CL &=& \frac{K_{Signal}}{\sqrt{K_{Signal} + K_{Bkgrnd}}}\\
\end{eqnarray*}

\subsubsection{Confidence Level of Signal in dataset - A at 14 TeV collision energy}\label{possiblecase} 
\begin{figure}[h!]
\begin{subfigure}{0.5\textwidth}
\begin{tikzpicture}
\hskip1.0cm \node (img1)  {\includegraphics[scale=0.42, trim={4.0cm 11.3cm 0 8.5cm}, clip]{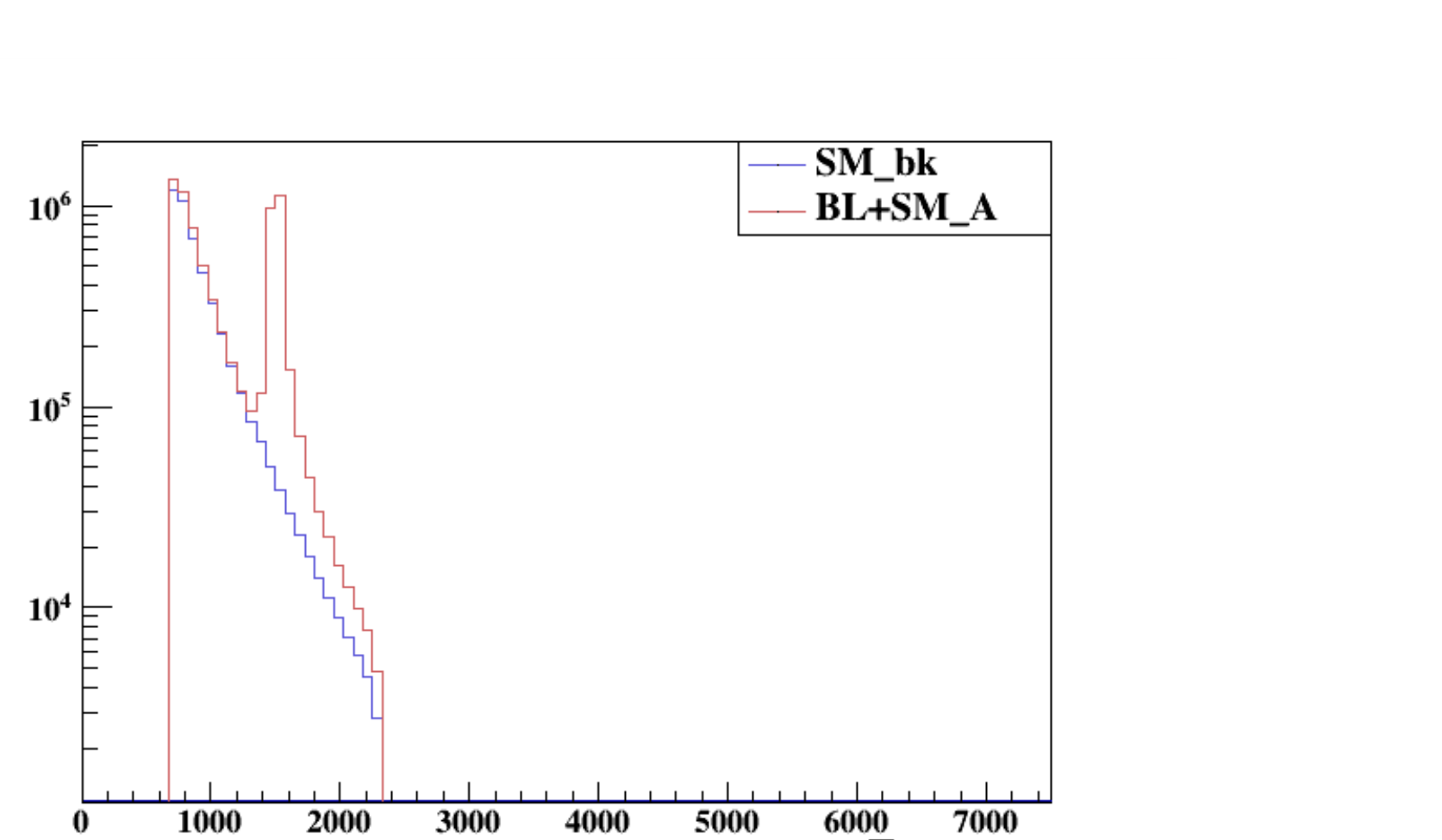}};
  \node[below=of img1, node distance=0cm, yshift=1cm,font=\color{black}] {$M_{{\mu}^+ \ {\mu}^-}$ (in GeV) };
  \node[right=of img1, node distance=0cm, rotate=90, anchor=center,yshift=8.4cm,font=\color{black}] {Evnts (300 $fb^{-1}$)};
\end{tikzpicture}
\end{subfigure}
\caption{Events distributions for datasets - A $\&$ SM at 14 TeV after cuts on $P_T$, $\eta$, $\&$ $M$ with $Z'_{B-L}$ signal with CL of 9.5$\sigma$ at $M_{{\mu}^+ \ {\mu}^-}$ = 1.5 TeV ($M_{Z'}^{dtaset-A}$) $\&$ $g'_1$ = 0.2.}
\label{fig:A14}
\end{figure}

\textcolor{white}{yxyzx}
\newline \\
We have studied the confidence level of $Z'_{B-L}$ signal of the dataset - A at 14 TeV with an integrated luminosity of $\mathcal{L}_{int}$ = 300 $fb^{-1}$.  The events distributions with the invariant mass for datasets - A $\&$ SM at 14 TeV CM energy are shown in Fig.~\ref{fig:A14}.  The number of events selected for Signal-Background analysis after the application of three successive kinematic cuts on the datasets that are just discussed, in the same order is listed in Table ~\ref{tab:Events}.  
\textbf{ }
\begin{table}[h]
  \begin{center}
  \renewcommand{\arraystretch}{0.3} 
\scalebox{0.95}{\begin{tabular}{|p{30.0mm}|p{20.0mm}|p{20.0mm}|}      \hline
     \vskip 0.2cm \cellcolor{Gray}  \textcolor{black}{Data Set - A \newline (Signal + Background)}  & \vskip 0.2cm \cellcolor{Gray} \textcolor{black}{Events \newline Retained \newline  K $ \ \pm \ \ \delta$K}   & \vskip 0.2cm \cellcolor{Gray} \textcolor{black}{Events\newline Rejected\newline R $\pm \ \delta$R } \\
     \hline \hline
  \vskip 0.2cm    \cellcolor{white} No cut \vskip 0.2cm &  \vskip 0.2cm  \cellcolor{white} 35351789 $\pm$ 10184.5 \vskip 0.2cm &  \vskip 0.2cm  \cellcolor{white} nil \vskip 0.2cm\\ 
      \hline
     \vskip 0.2cm \cellcolor{white} $P_T$ > 20.0 GeV \vskip 0.2cm  &  \vskip 0.2cm \cellcolor{white}  26308333 $\pm$ 9352 \vskip 0.2cm &  \vskip 0.2cm \cellcolor{white}  9043456 $\pm$ 4033   \vskip 0.2cm \\
    \hline  
    \vskip 0.2cm \cellcolor{white}  $\eta$ < 2.4 \vskip 0.2cm & \vskip 0.2cm \cellcolor{white} 26294016 $\pm$ 9348 \vskip 0.2cm & \vskip 0.2cm \cellcolor{white} 14317 $\pm$ 119 \vskip 0.2cm \\
      \hline
     \vskip 0.2cm \cellcolor{white} 692.7 GeV < M < 2307.3 GeV \vskip 0.2cm & \vskip 0.2cm \cellcolor{white} 1126.2 $\pm$ 33.6 \vskip 0.2cm & \vskip 0.2cm \cellcolor{white} 26292890 $\pm$ 9347 \vskip 0.2cm \\
    \hline
\vskip 0.2cm \cellcolor{Gray}  \textcolor{black}{Data Set - SM (Background Alone)}  & \vskip 0.2cm \cellcolor{Gray} \textcolor{black}{Events \newline Retained \newline  K $\pm \ \delta$K}   & \vskip 0.2cm \cellcolor{Gray} \textcolor{black}{Events\newline Rejected\newline R $\pm \ \delta$R } \\
     \hline \hline
  \vskip 0.2cm    \cellcolor{white} No cut \vskip 0.2cm &  \vskip 0.2cm  \cellcolor{white} 35527829 $\pm$ 10662.1 \vskip 0.2cm &  \vskip 0.2cm  \cellcolor{white} nil \vskip 0.2cm\\ 
      \hline
     \vskip 0.2cm \cellcolor{white} $P_T$ > 20.0 GeV \vskip 0.2cm  &  \vskip 0.2cm \cellcolor{white}  26949438 $\pm$ 9895 \vskip 0.2cm &  \vskip 0.2cm \cellcolor{white}  8578391 $\pm$ 3971  \vskip 0.2cm \\
    \hline  
    \vskip 0.2cm \cellcolor{white}  $\eta$ < 2.4 \vskip 0.2cm & \vskip 0.2cm \cellcolor{white} 26935139 $\pm$ 9890 \vskip 0.2cm & \vskip 0.2cm \cellcolor{white} 14298 $\pm$ 119 \vskip 0.2cm \\
      \hline
     \vskip 0.2cm \cellcolor{white} 692.7 GeV < M < 2307.3 GeV \vskip 0.2cm & \vskip 0.2cm \cellcolor{white} 807.6 $\pm$ 28.4 \vskip 0.2cm & \vskip 0.2cm \cellcolor{white} 26934331 $\pm$ 9890 \vskip 0.2cm \\
    \hline
    \end{tabular}}
    \caption{Selected and Rejected Events after the kinematic cuts on final state muon pairs for the datasets - A $\&$ SM at 14 TeV p-p Collisions }
  \label{tab:Events}
  \end{center}
\end{table}

\newpage
\indent The CL of $Z'_{B-L}$ signal has been calculated to be 9$\sigma$ which accounts for a possible experimental discovery.

\subsubsection{Comparison of datasets - A $\&$ SM at 7 and 8 TeV collisions with respective ATLAS Results}\label{impcase} 
\begin{figure}[h!]
\begin{subfigure}{0.48\textwidth}
\begin{tikzpicture}
\hskip1.0cm \node (img1)  {\includegraphics[scale=0.44, trim={4.0cm 11.3cm 0 8.5cm}, clip]{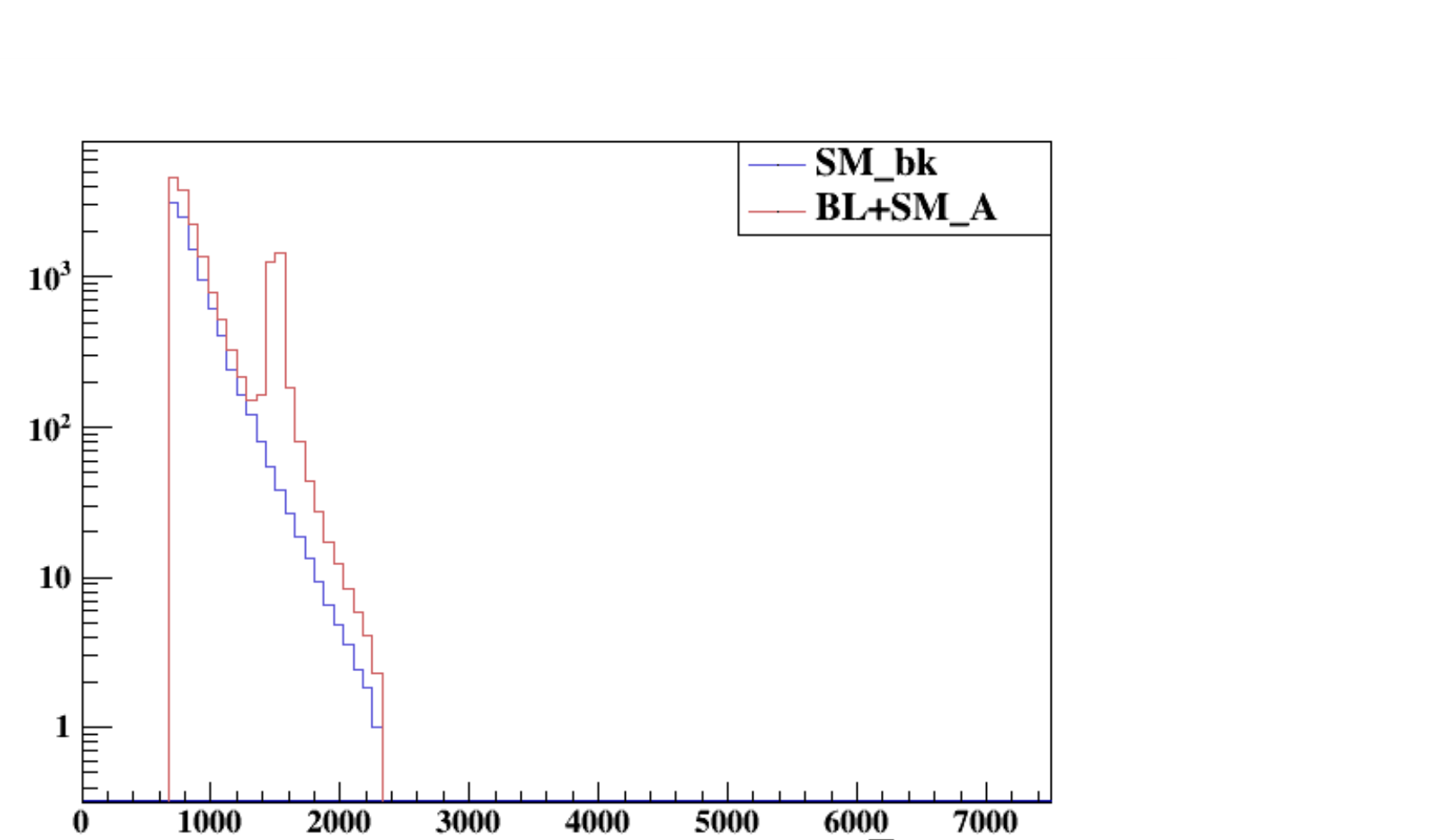}};
  \node[below=of img1, node distance=0cm, yshift=1cm,font=\color{black}] {$M_{{\mu}^+ \ {\mu}^-}$ (in GeV) };
  \node[right=of img1, node distance=0cm, rotate=90, anchor=center,yshift=8.8cm,font=\color{black}] {Evts (@ 1.21 $fb^{-1}$)};
\end{tikzpicture}
\caption{Events distribution of datasets - A $\&$ SM at 7 TeV collisions with $Z'_{B-L}$ signal of 0.1$\sigma$ significance.}
\label{fig:A7}
\end{subfigure}
\begin{subfigure}{0.48\textwidth}
\centering 
\includegraphics[scale=0.31,  trim={0 0.5cm 0 12.5cm}, clip]{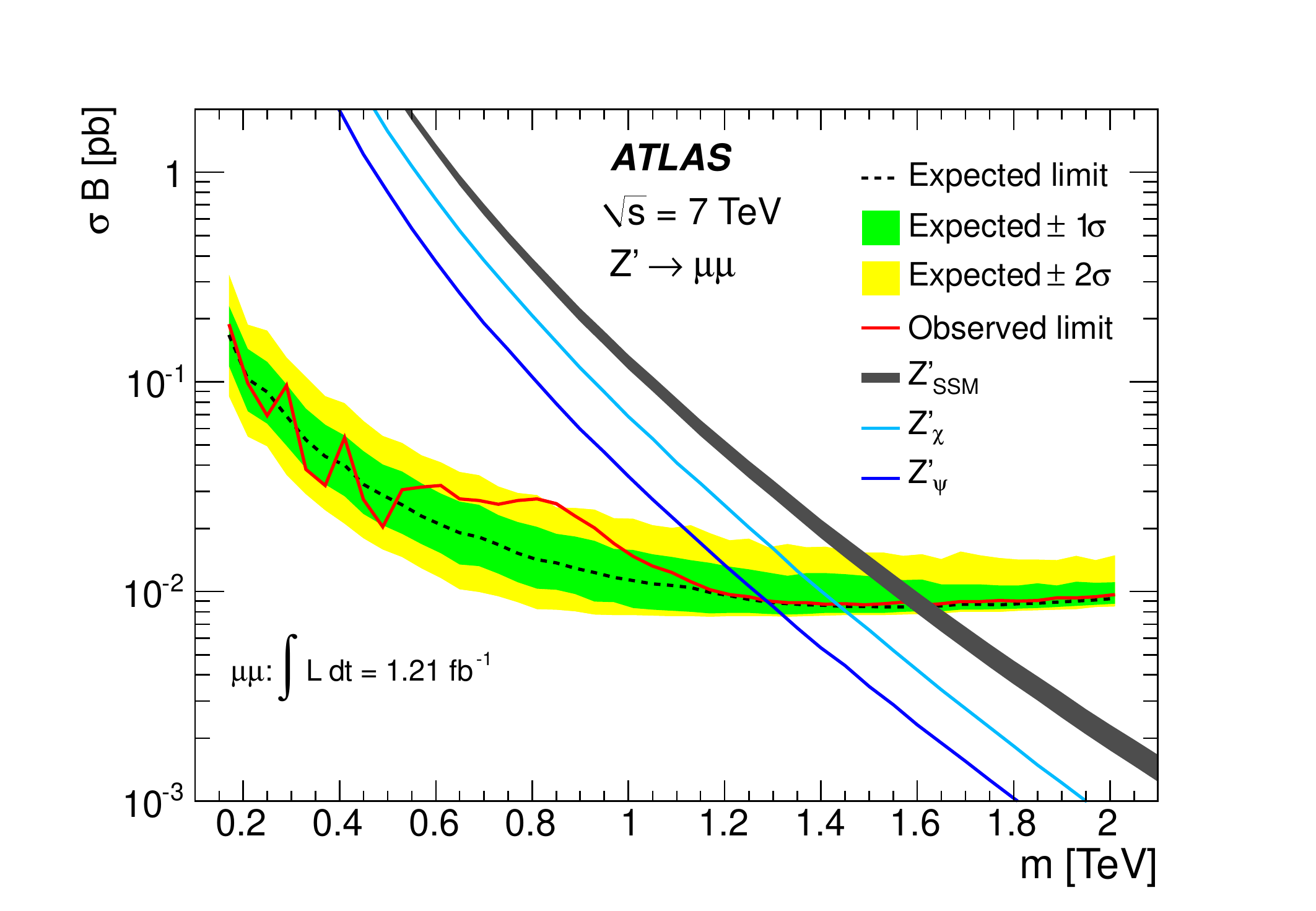}
\caption{Variation of cross section with invariant mass - ATLAS searches on $Z'$ in dimuon production channel at 7 TeV \cite{Collaboration:2011dca} Collisions with an integrated luminosity ($\mathcal{L}_{int}$) of 1.21 $fb^{-1}$ with no singularity at $M_{{\mu}^+ \ {\mu}^-}$ = 1.5 TeV.}
\label{fig:A7b}
\end{subfigure}
\caption{Comparison between datasets - A $\&$ SM with ATLAS searches on the neutral heavy resonance at 7 TeV p-p collisions with an integrated luminosity ($\mathcal{L}_{int}$) of 1.21 $fb^{-1}$.}
\label{fig:A7A7b}
\end{figure}

\textcolor{white}{xyz}\\ \\
The dataset - A studied at 7 TeV collision energy has been compared with the ATLAS experimental bounds on the searches of $Z'$ in the dimuon channel at 7 TeV LHC collisions with an integrated luminosity ($\mathcal{L}_{int}$) of 1.21 $fb^{-1}$ \cite{Collaboration:2011dca}.  The CL of the signal in the dataset - A, at 7 TeV collision is  0.1$\sigma$, which is merely a statistical fluctuation as shown in Fig.~\ref{fig:A7}.  We have confirmed with the ATLAS results of 7 TeV p-p collisions at 1.5 TeV where the cross section is continuous with no signal-peak, corresponding to the solid red line, titled ``Observed limit'' in Fig.~\ref{fig:A7b}.

\begin{figure}[h!]
\begin{subfigure}{0.48\textwidth}
\begin{tikzpicture}
\hskip1.0cm \node (img1)  {\includegraphics[scale=0.44, trim={4.0cm 11.3cm 0 8.5cm}, clip]{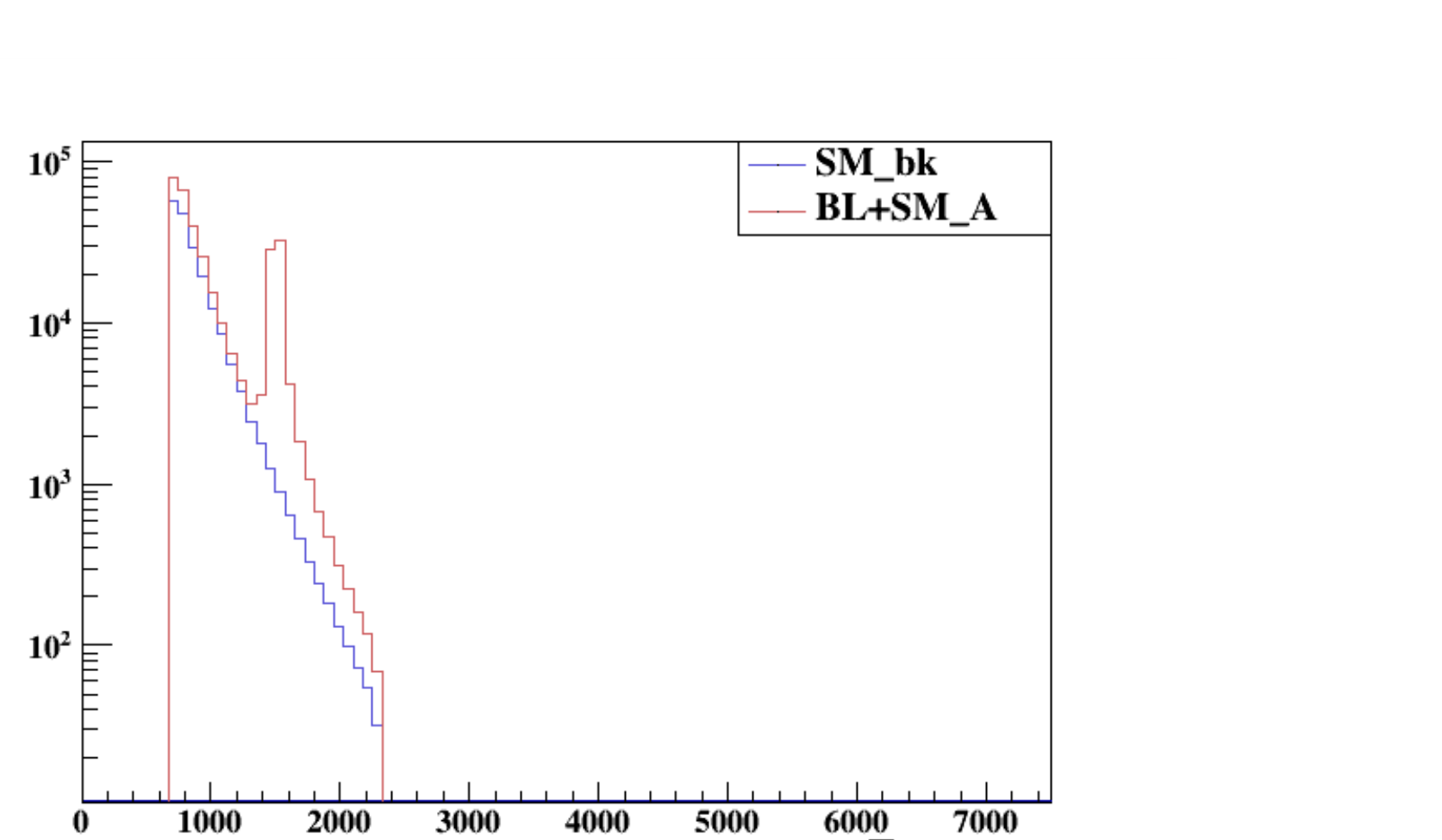}};
  \node[below=of img1, node distance=0cm, yshift=1cm,font=\color{black}] {$M_{{\mu}^+ \ {\mu}^-}$ (in GeV) };
  \node[right=of img1, node distance=0cm, rotate=90, anchor=center,yshift=9.0cm,font=\color{black}] {Evts (@ 20.5 $fb^{-1}$)};
\end{tikzpicture}
\caption{Events distribution for datasets - A $\&$ SM at 8 TeV collisions with the $Z'_{B-L}$ signal of 0.8$\sigma$ significance.}
\label{fig:A8a}
\end{subfigure}
\begin{subfigure}{0.48\textwidth}
\centering
\includegraphics[scale=0.3, clip]{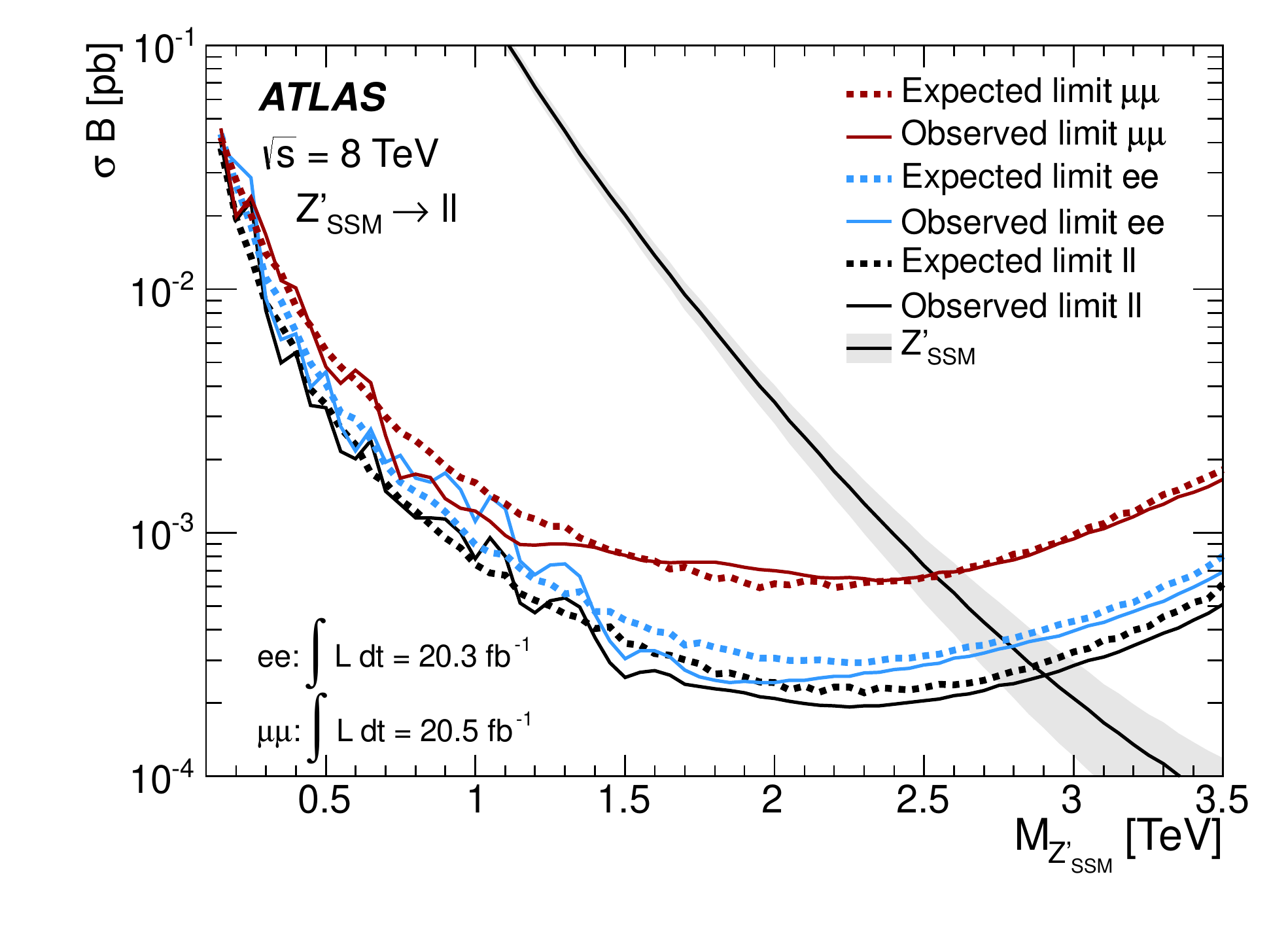}
\vskip0.1cm
\caption{Variation of cross section with invariant mass - ATLAS searches on $Z'$ in dimuon production channel at 8 TeV \cite{Aad:2014cka} Collisions with an integrated luminosity ($\mathcal{L}_{int}$) of 20.5 $fb^{-1}$ with no resonance at $M_{{\mu}^+ \ {\mu}^-}$ = 1.5 TeV. }
\label{fig:A8b}
\end{subfigure}
\caption{Comparison between datasets - A $\&$ SM with ATLAS searches on the neutral heavy resonance at 8 TeV p-p collisions with an integrated luminosity ($\mathcal{L}_{int}$) of 20.5 $fb^{-1}$.}
\label{fig:A8aA8b}
\end{figure}

A similar comparison between Figures~\ref{fig:A8a} $\&$~\ref{fig:A8b} has been made to confirm the $Z'_{B-L}$ resonance with CL of 0.8 $\sigma$ in the dataset - A, at 8 TeV p-p collision corresponds to no experimental possibility.  This detection impossibility is indicated by the continuous solid red line without any singularity, captioned with ``Observed limit ${\mu}{\mu}$'' in Fig.~\ref{fig:A8b} \cite{Aad:2014cka}.

\section{Conclusion}\label{conc}
Thus with the signal significance (CL) of 9$\sigma$ for dataset - A at 14 TeV CM energy with $\mathcal{L}_{int}$ = 300 $fb^{-1}$, this study predicts a potential discovery of the heavy neutral gauge boson ($Z'_{B-L}$) corresponding to the dataset - A with a mass $M_{Z'}$ of 1.5 TeV and a $Z'$ coupling strength $g'_1$ of 0.2 with the SM-fermions at the LHC as in Fig.~\ref{fig:A14}.

The comparative study between the datasets - A $\&$ SM with the ATLAS experimental searches on heavy resonances at 7 and 8 TeV on the dimuon channel confirms the non-observance of any $Z'$ boson.  And  this comparison validates our study of $Z'_{B-L}$ signal with CL of 0.1$\sigma$ and 0.8$\sigma$ at 7 and 8 TeV collisions of datasets - A $\&$ SM respectively as in Figs.~\ref{fig:A7A7b} $\&$~\ref{fig:A8aA8b}.

\appendix\label{app}
\section{\bf{Comparison of 7, 8 $\&$ 14 TeV simulated datasets}}
\begin{figure*}[bp!]
\begin{subfigure}{.6\columnwidth}
\begin{tikzpicture}
  \node (img1)  {\includegraphics[scale=0.29, trim={4.3cm 11.3cm 0cm 8.5cm}, clip]{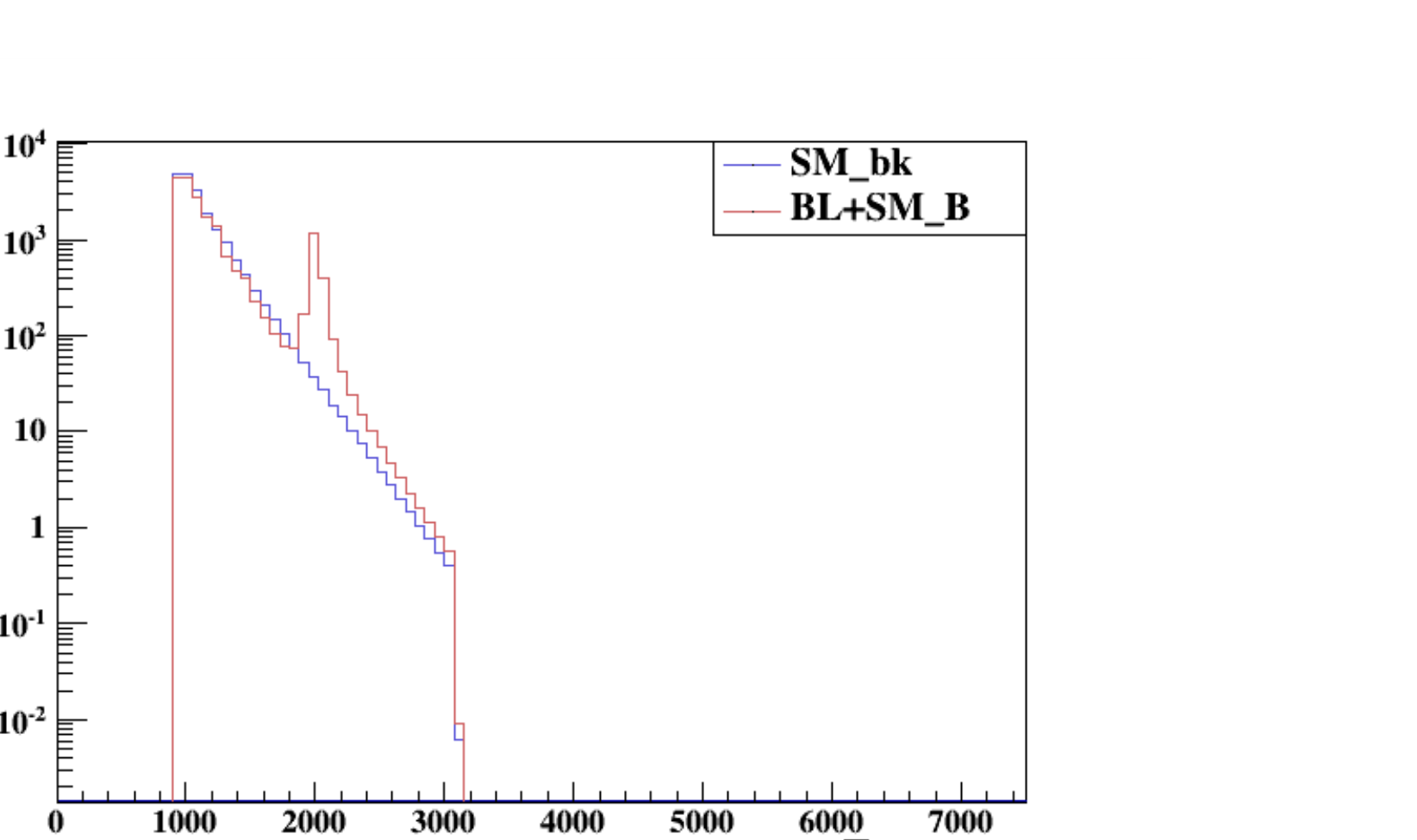}};
  \node[right=of img1, node distance=0cm, rotate=90, anchor=center,yshift=7.1cm,font=\color{black}] {Evts (1.21 $fb^{-1}$)};
\end{tikzpicture}
\caption{At $\sqrt{s}$ = 7 TeV with CL < 1$\sigma$}%
\label{subfiga}%
\end{subfigure}\hfill%
\begin{subfigure}{.6\columnwidth}
\begin{tikzpicture}
  \node (img1)  {\includegraphics[scale=0.29, trim={4.3cm 11.3cm 0cm 8.5cm}, clip]{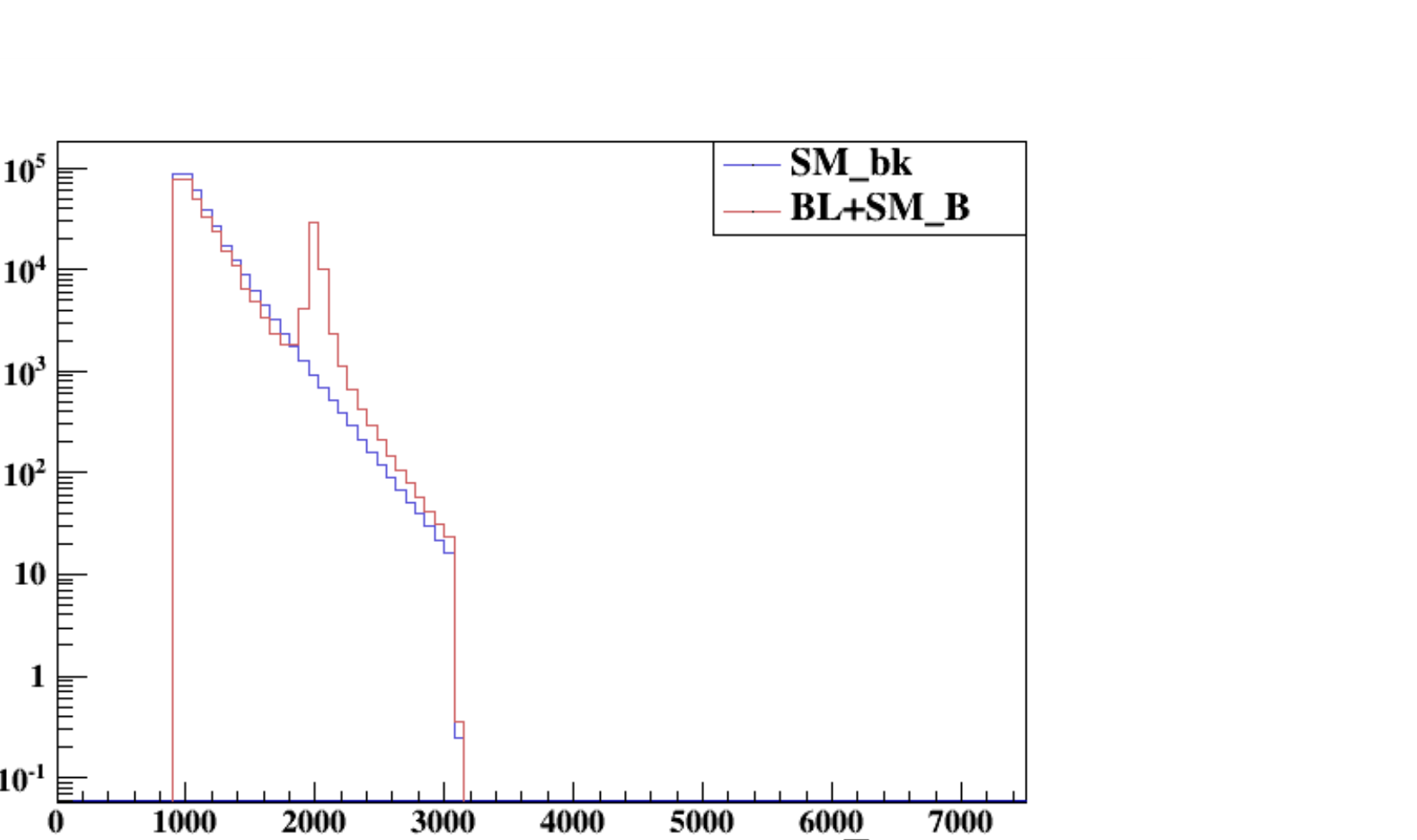}};
  \node[right=of img1, node distance=0cm, rotate=90, anchor=center,yshift=7.1cm,font=\color{black}] {Evts (20.5 $fb^{-1}$)};
\end{tikzpicture}
\caption{At $\sqrt{s}$ = 8 TeV with CL = 0.2$\sigma$}%
\label{subfigb}%
\end{subfigure}\hfill%
\begin{subfigure}{.6\columnwidth}
\begin{tikzpicture}
 \node (img1)  {\includegraphics[scale=0.29, trim={4.3cm 11.3cm 0cm 8.5cm}, clip]{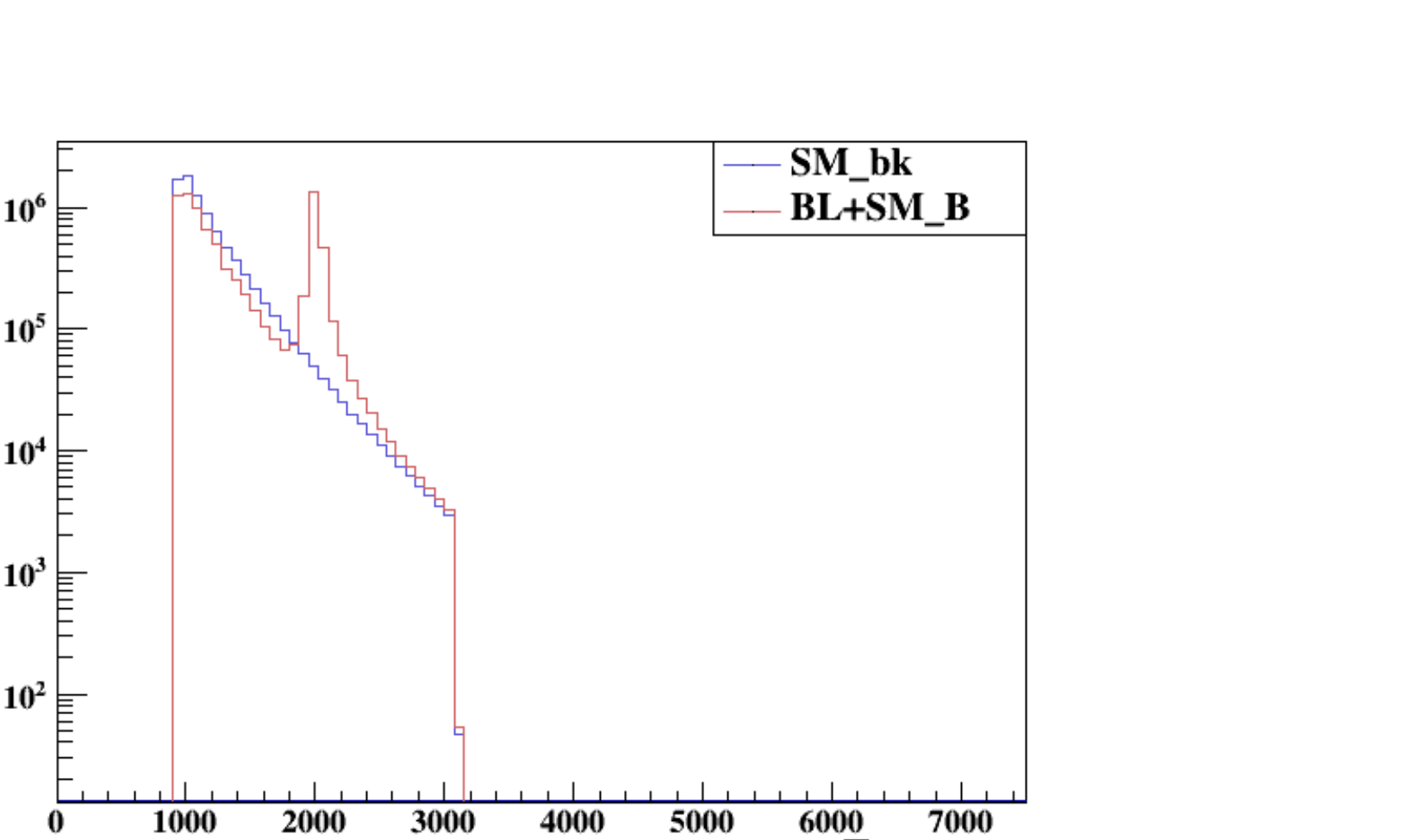}};
  \node[right=of img1, node distance=0cm, rotate=90, anchor=center,yshift=7.1cm,font=\color{black}] {Evts (300 $fb^{-1}$)};
\end{tikzpicture}
\caption{At $\sqrt{s}$ = 14 TeV with CL = 4.3$\sigma$}%
\label{subfigc}%
\end{subfigure}%
\caption{\textcolor{black}{Dataset - B} with cuts, $P_T$ \textsuperscript{\ref{note2}}, $\eta$ \textsuperscript{\ref{note2}} and \textcolor{blue}{923.6 GeV} $\ \leq\ \textcolor{blue}{M_{{\mu}^+ \ {\mu}^-}} \ \leq\ $ \textcolor{blue}{3076.4 GeV} ( for \textcolor{blue}{ ${\Gamma}_{SW}^{signl-B}$} ) \textsuperscript{\ref{note3}}}
\label{fig:B}
\begin{subfigure}{.6\columnwidth}
\begin{tikzpicture}
  \node (img1)  {\includegraphics[scale=0.29, trim={4.3cm 11.3cm 0cm 8.5cm}, clip]{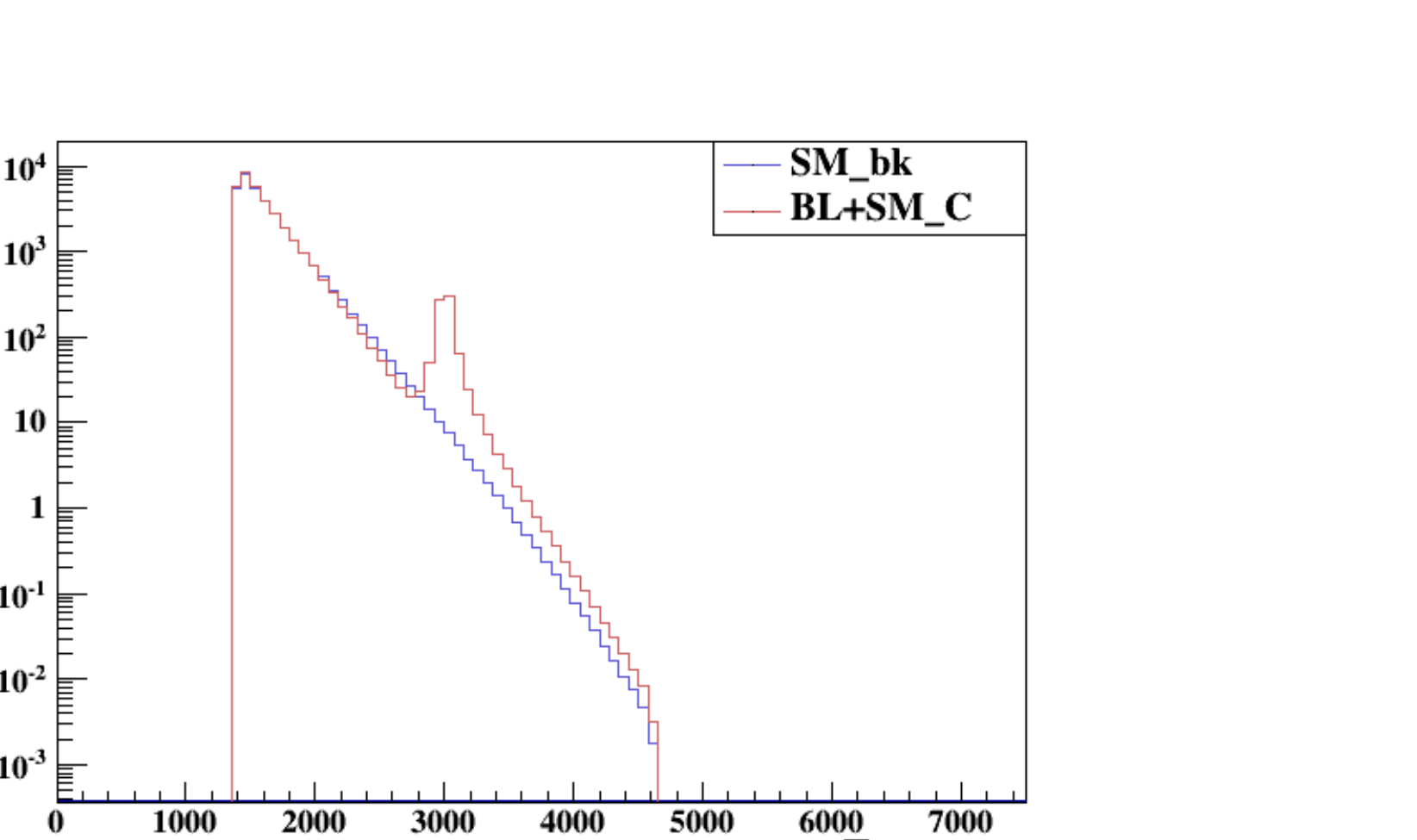}};
  \node[right=of img1, node distance=0cm, rotate=90, anchor=center,yshift=7.1cm,font=\color{black}] {Evts (1.21 $fb^{-1}$)};
\end{tikzpicture}
\caption{At $\sqrt{s}$ = 7 TeV with CL < 1$\sigma$}%
\label{subfiga}%
\end{subfigure}\hfill%
\begin{subfigure}{.6\columnwidth}
\begin{tikzpicture}
  \node (img1)  {\includegraphics[scale=0.29, trim={4.3cm 11.3cm 0cm 8.5cm}, clip]{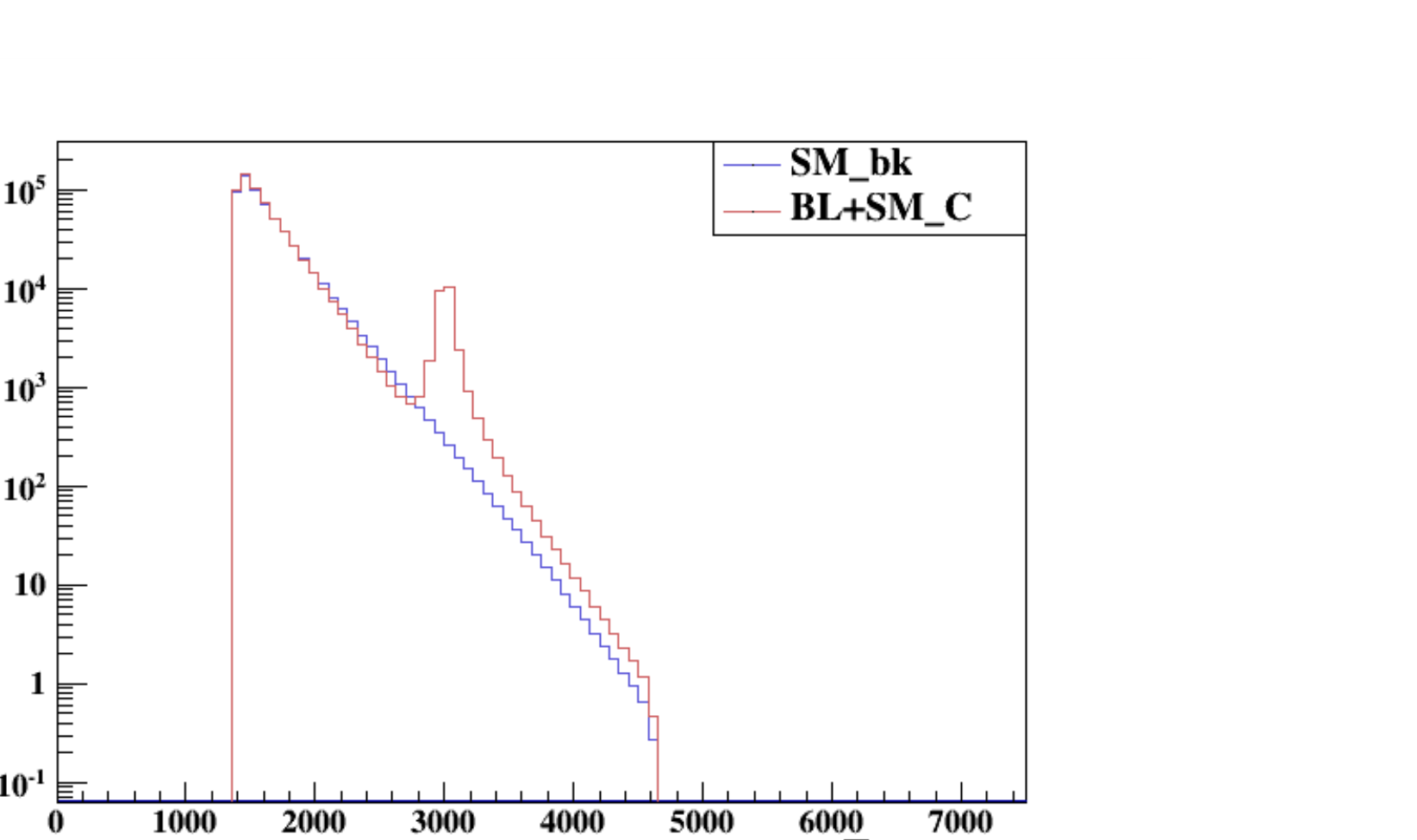}};
  \node[right=of img1, node distance=0cm, rotate=90, anchor=center,yshift=7.1cm,font=\color{black}] {Evts (20.5 $fb^{-1}$)};
\end{tikzpicture}
\caption{At $\sqrt{s}$ = 8 TeV with CL < < 1$\sigma$}%
\label{subfigb}%
\end{subfigure}\hfill%
\begin{subfigure}{.6\columnwidth}
\begin{tikzpicture}
 \node (img1)  {\includegraphics[scale=0.29, trim={4.3cm 11.3cm 0cm 8.5cm}, clip]{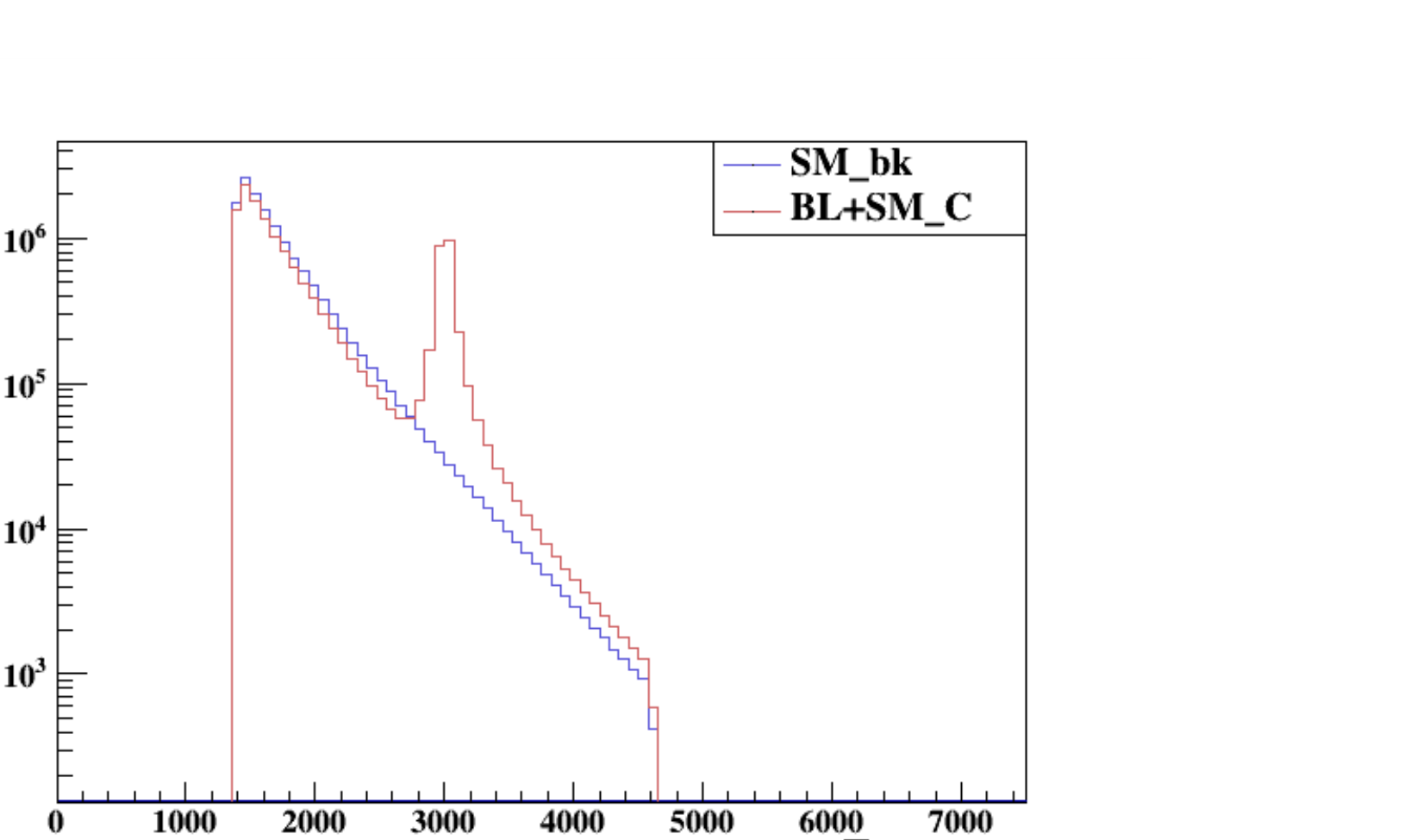}};
  \node[right=of img1, node distance=0cm, rotate=90, anchor=center,yshift=7.1cm,font=\color{black}] {Evts (300 $fb^{-1}$)};
\end{tikzpicture}
\caption{At $\sqrt{s}$ = 14 TeV with CL = 1$\sigma$}%
\label{subfigc}%
\end{subfigure}%
\caption{\textcolor{black}{Dataset - C} with cuts, $P_T$ \textsuperscript{\ref{note2}}, $\eta$ \textsuperscript{\ref{note2}} and \textcolor{blue}{1385.4 GeV} $\ \leq\ \textcolor{blue}{M_{{\mu}^+ \ {\mu}^-}} \ \leq\ $ \textcolor{blue}{4614.6 GeV} ( for \textcolor{blue}{ ${\Gamma}_{SW}^{signl-C}$} ) \textsuperscript{\ref{note3}}}
\label{fig:C}
\begin{subfigure}{.6\columnwidth}
\begin{tikzpicture}
  \node (img1)  {\includegraphics[scale=0.29, trim={4.3cm 11.3cm 0cm 8.5cm}, clip]{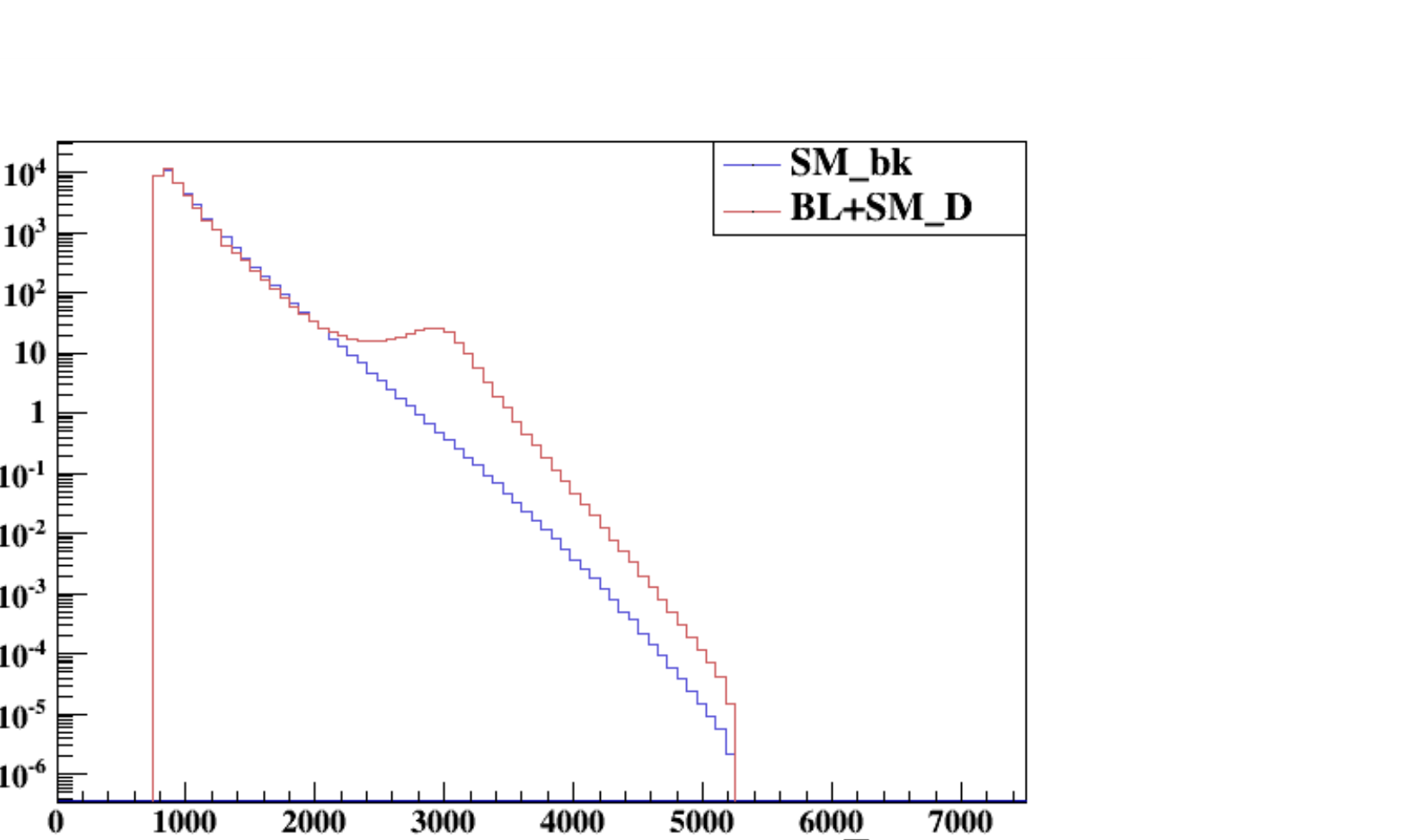}};
  \node[right=of img1, node distance=0cm, rotate=90, anchor=center,yshift=7.1cm,font=\color{black}] {Evts (1.21 $fb^{-1}$)};
\end{tikzpicture}
\caption{At $\sqrt{s}$ = 7 TeV with CL < 1$\sigma$}%
\label{subfiga}%
\end{subfigure}\hfill%
\begin{subfigure}{.6\columnwidth}
\begin{tikzpicture}
  \node (img1)  {\includegraphics[scale=0.29, trim={4.3cm 11.3cm 0cm 8.5cm}, clip]{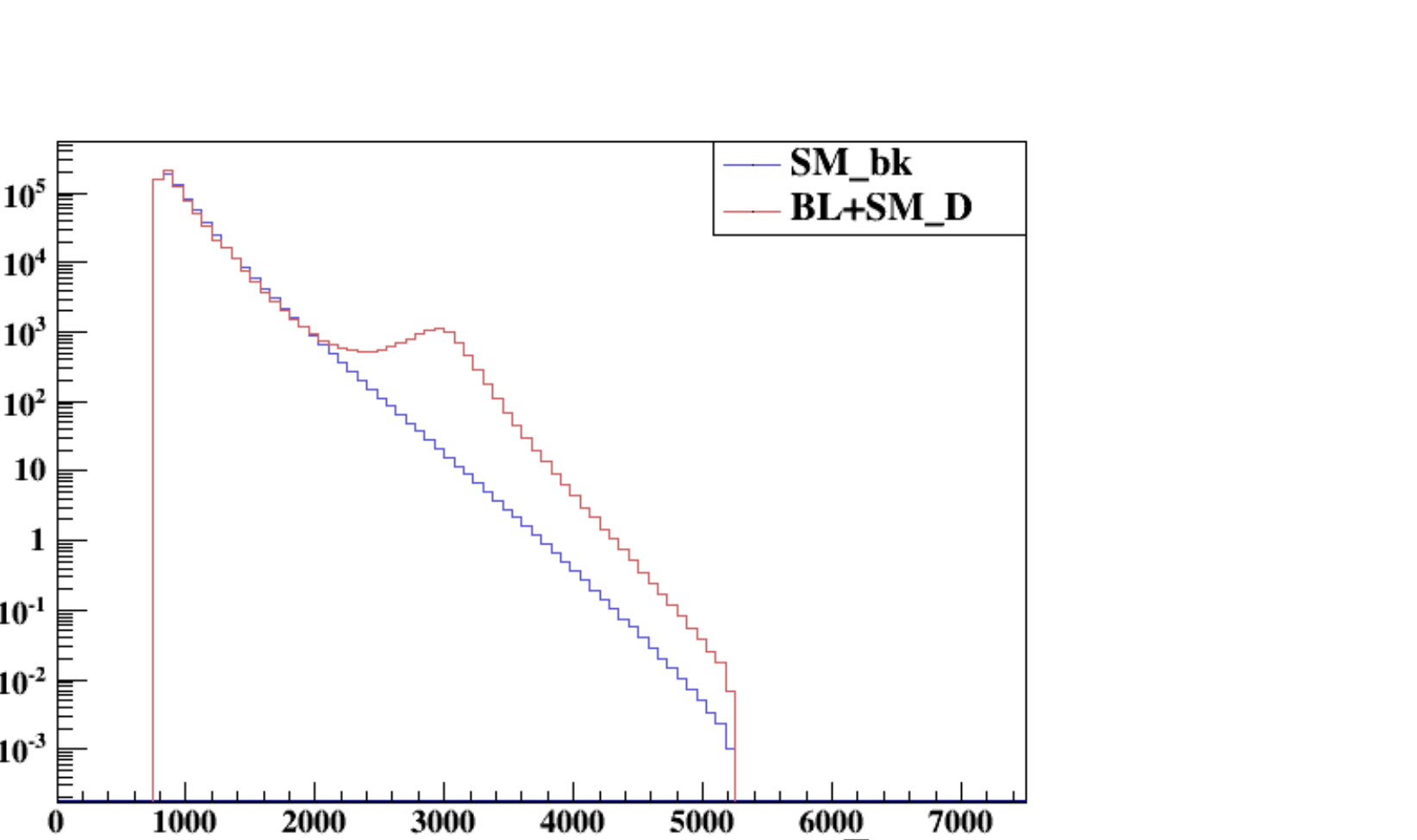}};
  \node[right=of img1, node distance=0cm, rotate=90, anchor=center,yshift=7.1cm,font=\color{black}] {Evts (20.5 $fb^{-1}$)};
\end{tikzpicture}
\caption{At $\sqrt{s}$ = 8 TeV with CL < < 1$\sigma$}%
\label{subfigb}%
\end{subfigure}\hfill%
\begin{subfigure}{.6\columnwidth}
\begin{tikzpicture}
 \node (img1)  {\includegraphics[scale=0.29, trim={4.3cm 11.3cm 0cm 8.5cm}, clip]{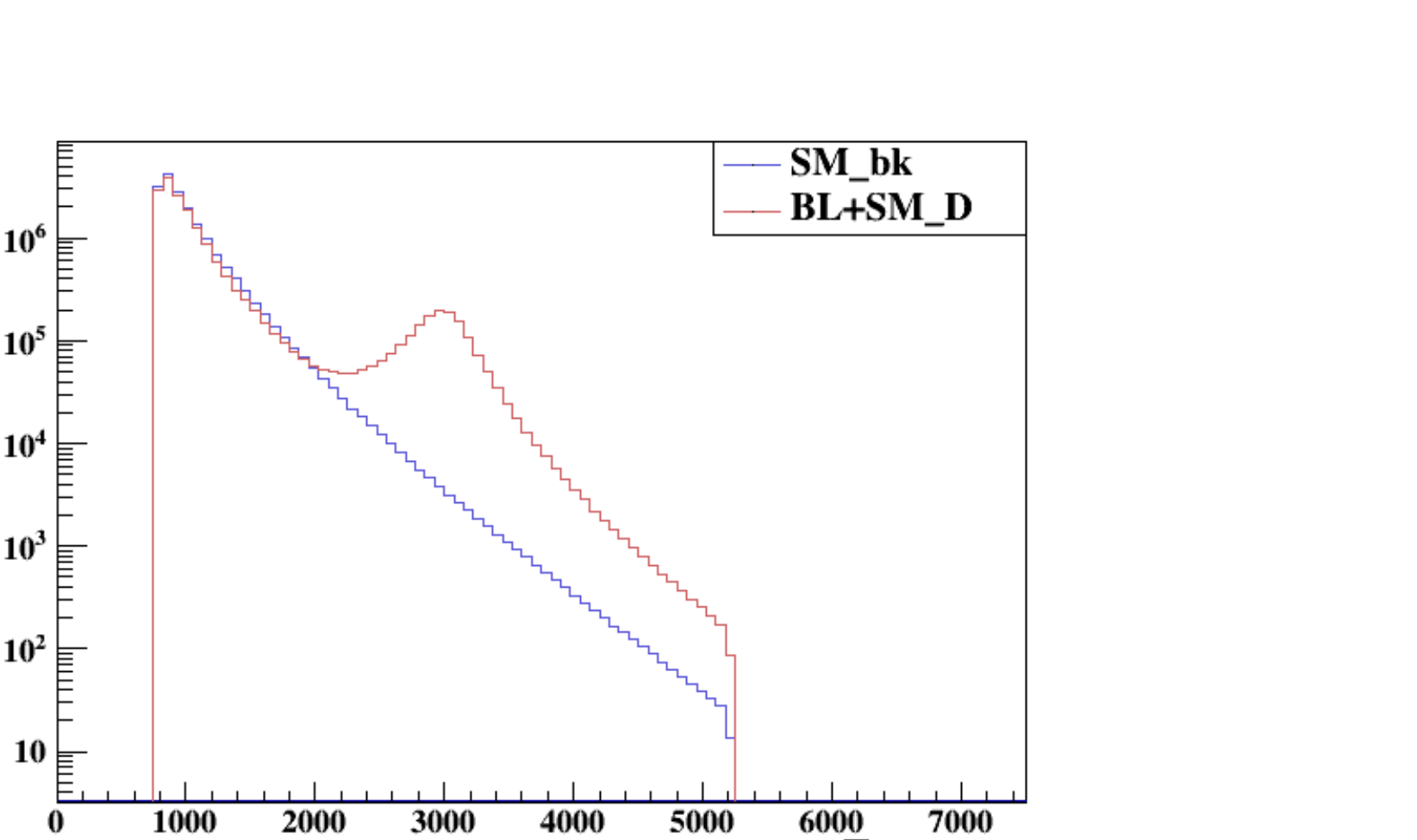}};
  \node[right=of img1, node distance=0cm, rotate=90, anchor=center,yshift=7.1cm,font=\color{black}] {Evts (300 $fb^{-1}$)};
\end{tikzpicture}
\caption{At $\sqrt{s}$ = 14 TeV with CL = 0.7$\sigma$}%
\label{subfigc}%
\end{subfigure}%
\caption{\textcolor{black}{Dataset - D} with cuts, $P_T$ \textsuperscript{\ref{note2}}, $\eta$ \textsuperscript{\ref{note2}} and \textcolor{blue}{783.8 GeV} $\ \leq\ \textcolor{blue}{M_{{\mu}^+ \ {\mu}^-}} \ \leq\ $ \textcolor{blue}{5216.2 GeV} ( for \textcolor{blue}{ ${\Gamma}_{SW}^{signl-D}$} ) \textsuperscript{\ref{note3}}}
\label{fig:D}
\begin{subfigure}{.6\columnwidth}
\begin{tikzpicture}
  \node (img1)  {\includegraphics[scale=0.29, trim={4.3cm 11.3cm 0cm 8.5cm}, clip]{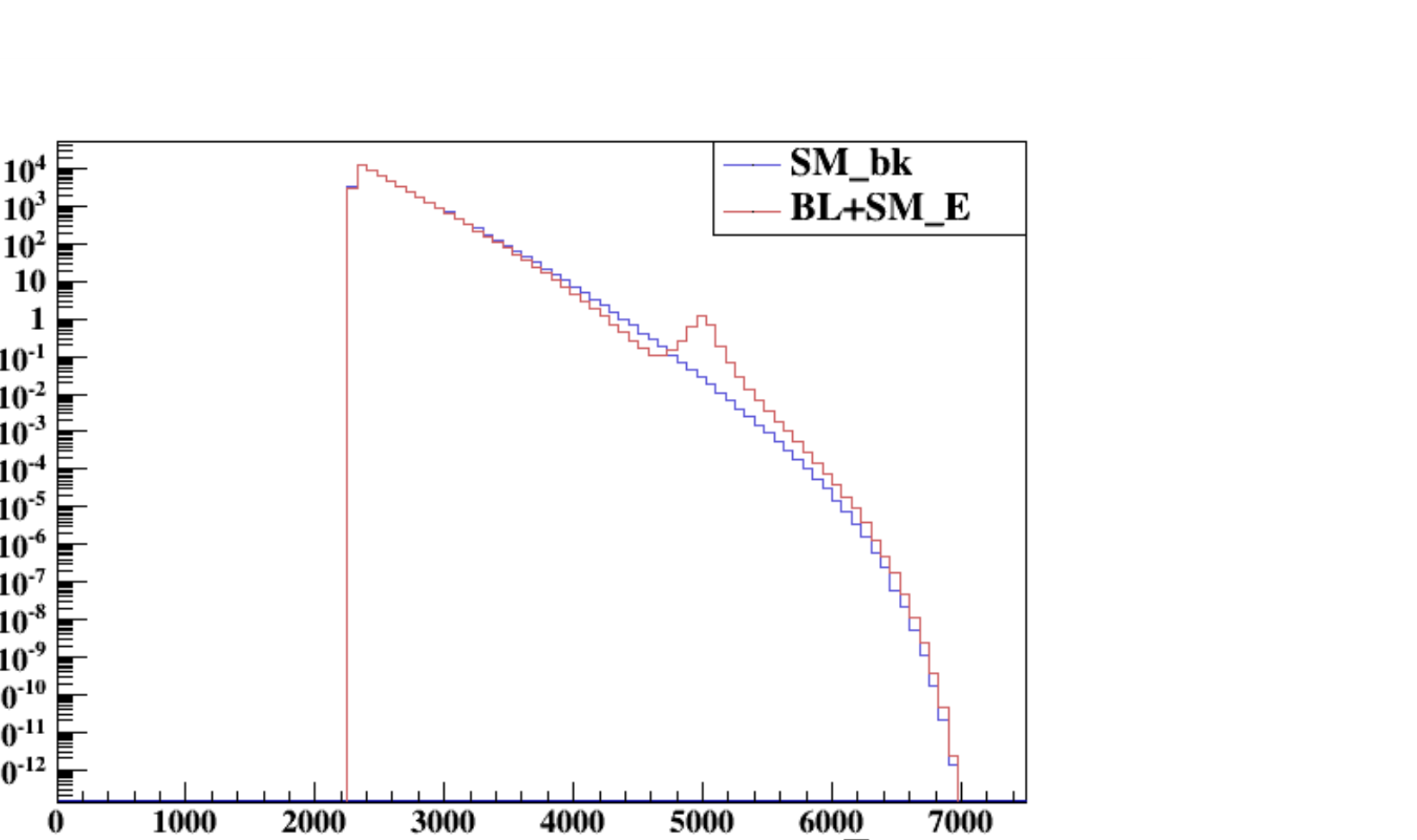}};
  \node[right=of img1, node distance=0cm, rotate=90, anchor=center,yshift=7.1cm,font=\color{black}] {Evts (1.21 $fb^{-1}$)};
\end{tikzpicture}
\caption{At $\sqrt{s}$ = 7 TeV with CL < 1$\sigma$}%
\label{subfiga}%
\end{subfigure}\hfill%
\begin{subfigure}{.6\columnwidth}
\begin{tikzpicture}
  \node (img1)  {\includegraphics[scale=0.29, trim={4.3cm 11.3cm 0cm 8.5cm}, clip]{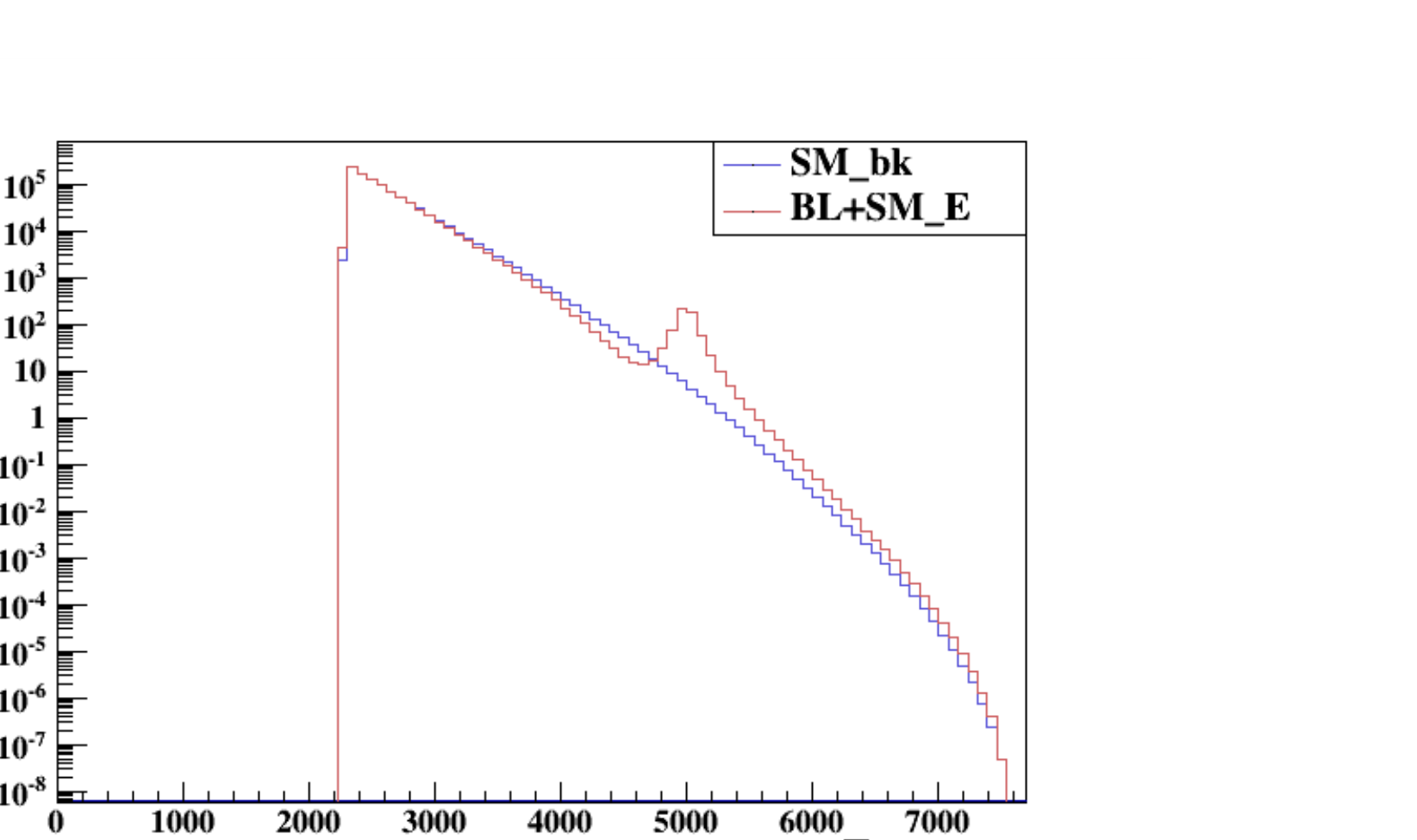}};
  \node[right=of img1, node distance=0cm, rotate=90, anchor=center,yshift=7.1cm,font=\color{black}] {Evts (20.5 $fb^{-1}$)};
\end{tikzpicture}
\caption{At $\sqrt{s}$ = 8 TeV with CL < < 1$\sigma$}%
\label{subfigb}%
\end{subfigure}\hfill%
\begin{subfigure}{.6\columnwidth}
\begin{tikzpicture}
 \node (img1)  {\includegraphics[scale=0.29, trim={4.3cm 11.3cm 0cm 8.5cm}, clip]{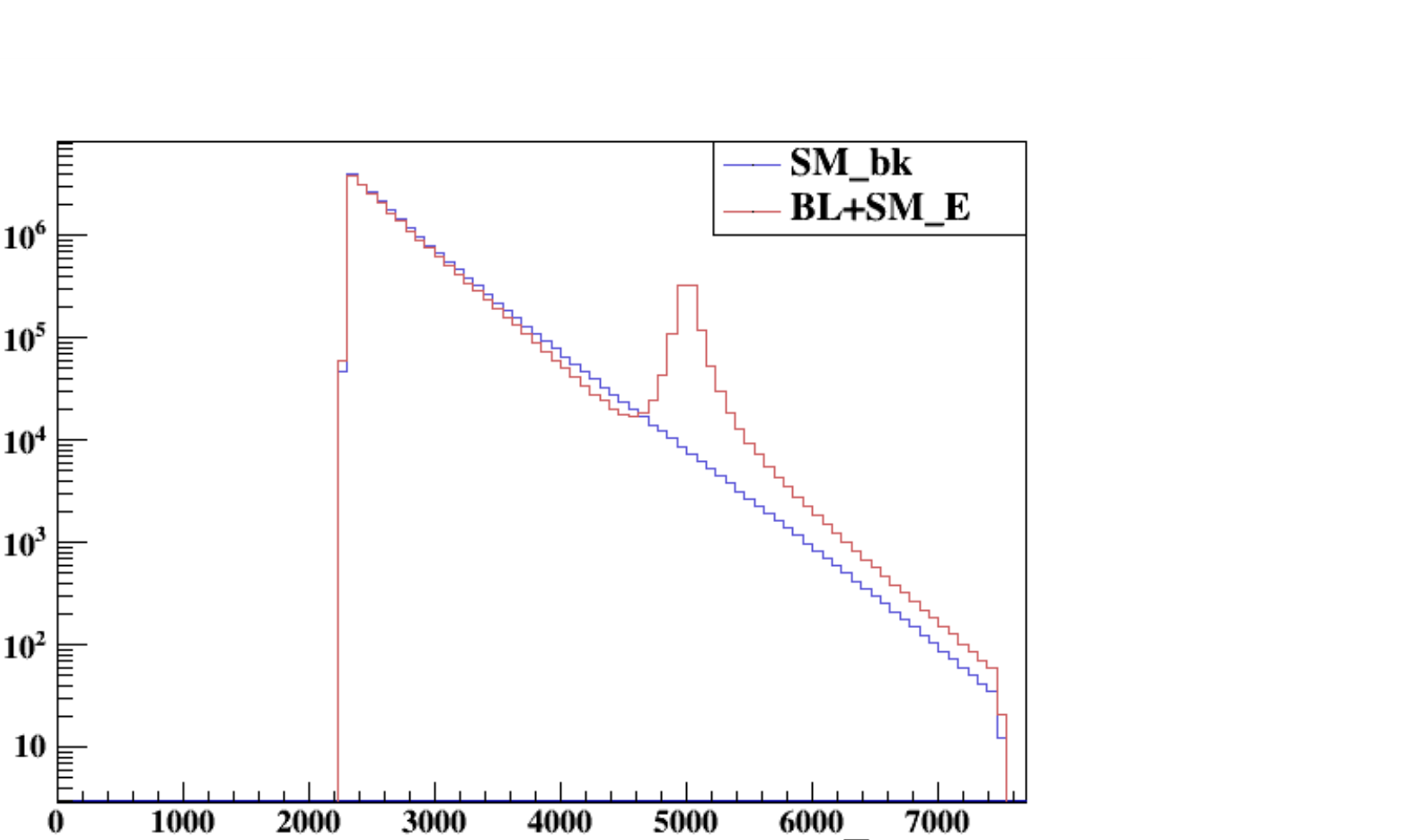}};
  \node[right=of img1, node distance=0cm, rotate=90, anchor=center,yshift=7.1cm,font=\color{black}] {Evts (300 $fb^{-1}$)};
\end{tikzpicture}
\caption{At $\sqrt{s}$ = 14 TeV with CL = 0$\sigma$}%
\label{subfigc}%
\end{subfigure}%
\caption{\textcolor{black}{Dataset - E} with cuts, $P_T$ \textsuperscript{\ref{note2}}, $\eta$ \textsuperscript{\ref{note2}} and \textcolor{blue}{2309.0 GeV} $\ \leq\ \textcolor{blue}{M_{{\mu}^+ \ {\mu}^-}} \ \leq\ $ \textcolor{blue}{7691.0 GeV} ( for \textcolor{blue}{ ${\Gamma}_{SW}^{signl-E}$} ) \textsuperscript{\ref{note3}}}
\label{fig:E}
\begin{subfigure}{.6\columnwidth}
\begin{tikzpicture}
  \node (img1)  {\includegraphics[scale=0.29, trim={4.3cm 11.3cm 0cm 8.5cm}, clip]{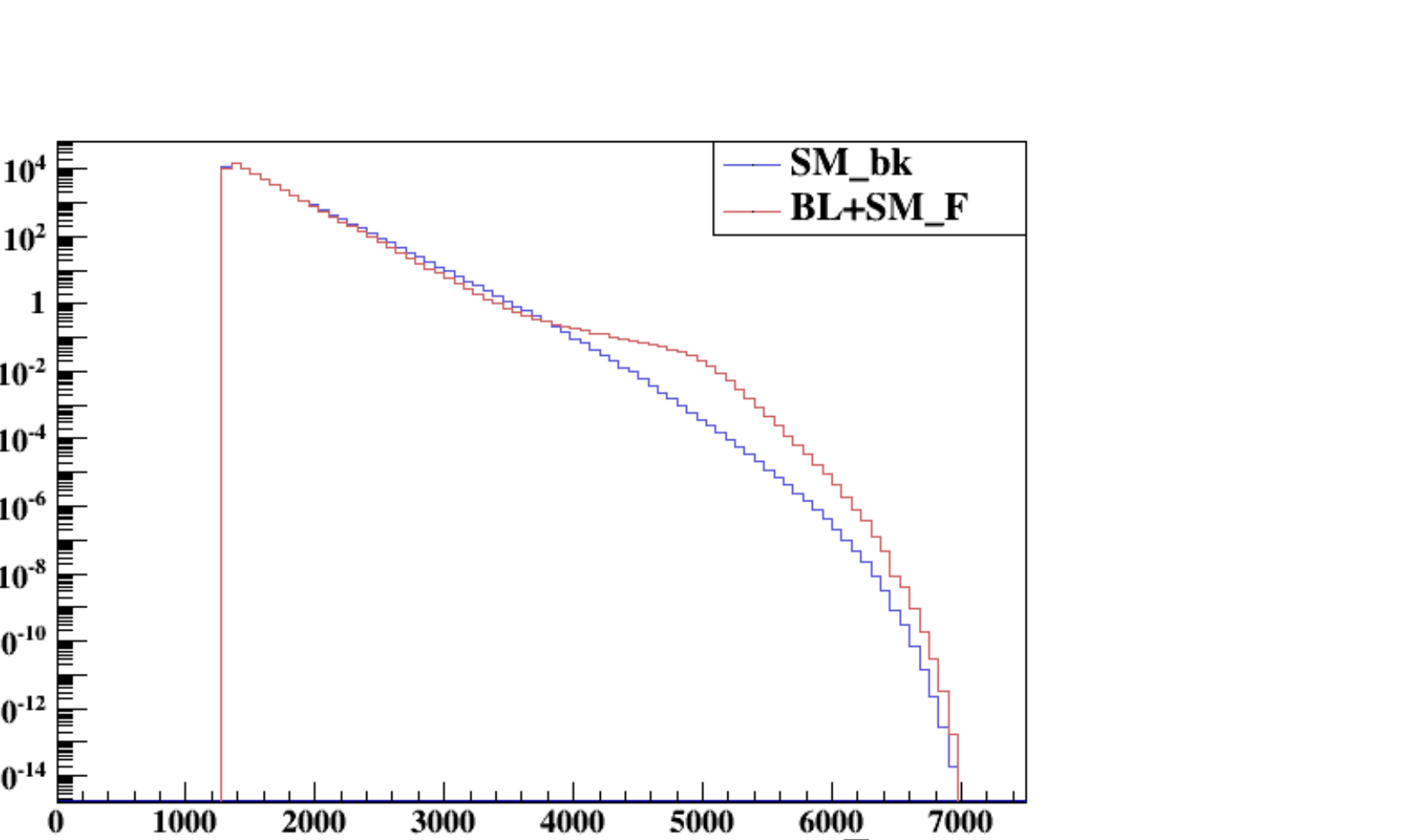}};
  \node[below=of img1, node distance=0cm, yshift=1cm,font=\color{black}] {M (in GeV) };
  \node[right=of img1, node distance=0cm, rotate=90, anchor=center,yshift=7.1cm,font=\color{black}] {Evts (1.21 $fb^{-1}$)};
\end{tikzpicture}
\caption{At $\sqrt{s}$ = 7 TeV with CL < 1$\sigma$}%
\label{subfiga}%
\end{subfigure}\hfill%
\begin{subfigure}{.6\columnwidth}
\begin{tikzpicture}
  \node (img1)  {\includegraphics[scale=0.29, trim={4.3cm 11.3cm 0cm 8.5cm}, clip]{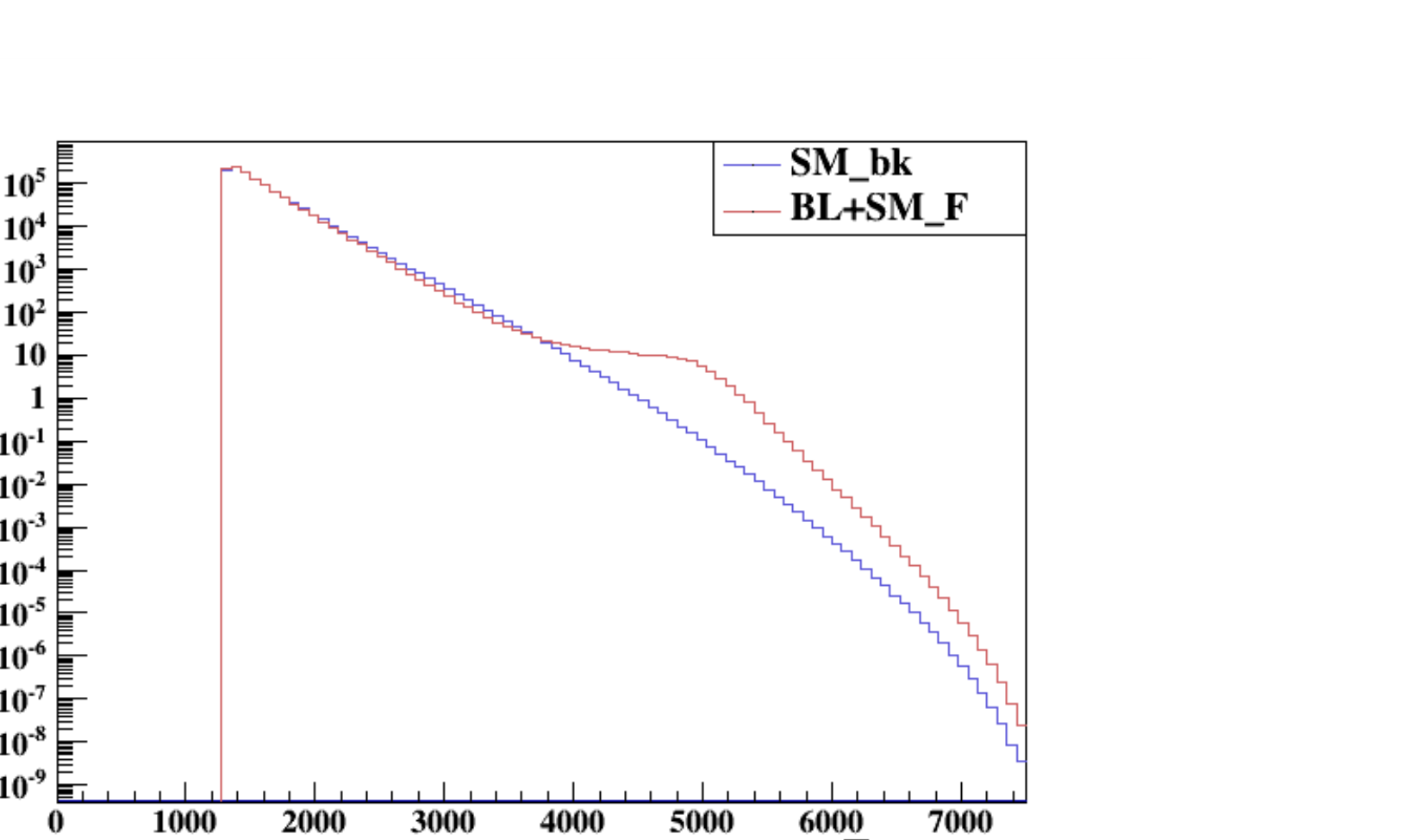}};
  \node[below=of img1, node distance=0cm, yshift=1cm,font=\color{black}] {M (in GeV) };
  \node[right=of img1, node distance=0cm, rotate=90, anchor=center,yshift=7.1cm,font=\color{black}] {Evts (20.5 $fb^{-1}$)};
\end{tikzpicture}
\caption{At $\sqrt{s}$ = 8 TeV with CL < < 1$\sigma$}%
\label{subfigb}%
\end{subfigure}\hfill%
\begin{subfigure}{.6\columnwidth}
\begin{tikzpicture}
 \node (img1)  {\includegraphics[scale=0.29, trim={4.3cm 11.3cm 0cm 8.5cm}, clip]{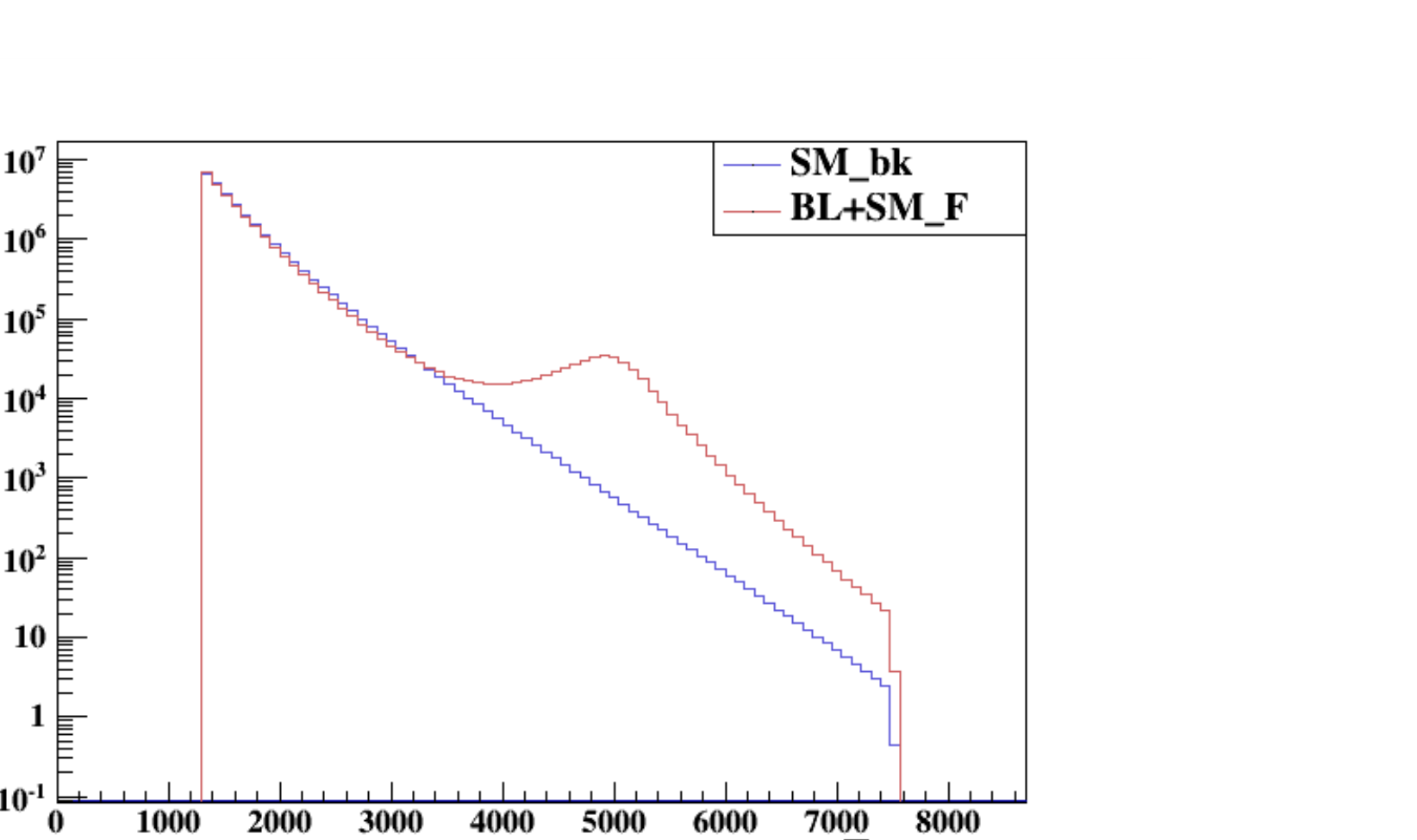}};
  \node[below=of img1, node distance=0cm, yshift=1cm,font=\color{black}] {M (in GeV) };
  \node[right=of img1, node distance=0cm, rotate=90, anchor=center,yshift=7.1cm,font=\color{black}] {Evts (300 $fb^{-1}$)};
\end{tikzpicture}
\caption{At $\sqrt{s}$ = 14 TeV with CL < < 1$\sigma$}%
\label{subfigc}%
\end{subfigure}%
\caption{\textcolor{black}{Dataset - F} with cuts, $P_T$ \textsuperscript{\ref{note2}}, $\eta$ \textsuperscript{\ref{note2}} and \textcolor{blue}{1306.4 GeV} $\ \leq\ \textcolor{blue}{M_{{\mu}^+ \ {\mu}^-}} \ \leq\ $ \textcolor{blue}{8693.6 GeV} ( for \textcolor{blue}{ ${\Gamma}_{SW}^{signl-F}$} ) \textsuperscript{\ref{note3}}}
\label{fig:F}
\end{figure*}

The event distribution of the remaining datasets with Muon Pairs' Invariant mass, studied at 7, 8 $\&$ 14 TeV are compared in this appendix (Figs.~\ref{fig:B},~\ref{fig:C},~\ref{fig:D},~\ref{fig:E} $\&$~\ref{fig:F} corresponding to datasets - B, C, D, E, $\&$ F respectively).  In all these cases, the CL of $Z'_{B-L}$ signal is less than 3$\sigma$, which are accounted for no experimental discoveries.
 
\section*{Acknowledgement}
This study of $Z'_{B-L}$ was carried out as a dissertation thesis of M.Sc. Particle Physics program, under the guidance of Dr Sebastian J{\"a}ger, University of Sussex, U.K.  I express my sincere thanks and gratitude to my thesis adviser here. \\ 
\indent I thank the Physical Research Laboratory (PRL), Ahmedabad for hospitality and support during which I got an opportunity to present our work in the Pheno01@\hskip15.0cm IISERMohali workshop.  My humble thanks to the organisers of Pheno01@IISERMohali workshop for having and supporting my stay at the IISER-Mohali, India. \\
\indent  The final part of this work in preparing the manuscr-\hskip 15.0cm ipt was done during my visiting tenure in Harish-Chand-\hskip 15.0cm ra Research Institute (HRI), India. I am extending my sincere gratitude to Prof. Biswarup Mukhopadhyaya, and the Regional Centre for Accelerator - based Particle Physics (RECAPP) - HRI for supporting me and my learning process.
 

\bibliographystyle{plainnat}


\end{document}